\documentstyle[12pt]{article}

\def\+{{+\!\!\!+}} 
\def\pp{\mbox{\tiny${}_{\stackrel\+ =}$}}

\def\d{\partial}

\def\th{\theta}

\def\P{\Phi}

\def\p{\psi} 
 
\def\e{\varepsilon}

\def\pmb#1{\setbox0=\hbox{#1}%
\kern.0em\copy0\kern-\wd0 
\kern-.04em\copy0\kern-\wd0 
\kern.08em\copy0\kern-\wd0 
\kern-.04em\raise.0433em\box0 }         
 
\def\rank{\textstyle{\rm{rank}}} 
 
\newcommand{\nc}{\newcommand} 
\nc{\beq}{\begin{equation}} 
\nc{\eeq}[1]{\label{#1}\end{equation}} 
\nc{\ber}{\begin{eqnarray}} 
\nc{\eer}[1]{\label{#1}\end{eqnarray}} 
\nc{\pek}[1]{\cite{#1}} 
\nc{\enr}[1]{(\ref{#1})} 
\nc{\kal}[1]{{\cal{#1}}} 
\nc{\dott}{\;\cdot\;} 
\newcommand{\Section}[1]{\section{#1} \setcounter{equation}{0}} 
 
\begin{document} 
\newcommand{\inv}[1]{{#1}^{-1}} 
\renewcommand{\theequation}{\thesection.\arabic{equation}} 
\newcommand{\be}{\begin{equation}} 
\newcommand{\ee}{\end{equation}} 
\newcommand{\bea}{\begin{eqnarray}} 
\newcommand{\eea}{\end{eqnarray}} 
\newcommand{\re}[1]{(\ref{#1})} 
\newcommand{\qv}{\quad ,} 
\newcommand{\qp}{\quad .} 
\begin{center} 
                                \hfill UUITP-11/02\\
                                \hfill   hep-th/0209098\\ 
\vskip .3in \noindent 
 
\vskip .1in 
 
{\large \bf{N=2 Boundary conditions for  
 non-linear sigma models and Landau-Ginzburg models}} 
\vskip .2in 
 
{\bf Ulf Lindstr\"om}$^a$\footnote{e-mail address: ul@physto.se} 
 and  {\bf Maxim Zabzine}$^{b}$\footnote{e-mail address: zabzine@fi.infn.it} \\ 
 
 
\vskip .15in 
 
\vskip .15in 
$^a${\em Department of Theoretical Physics, \\
Uppsala University,
Box 803, SE-751 08 Uppsala, Sweden }\\ 
\vskip .15in 
$^b${\em INFN Sezione di Firenze, Dipartimento di Fisica\\ 
 Via Sansone 1, I-50019 Sesto F.no (FI), Italy} 
 
\bigskip 
 
 
 \vskip .1in 
\end{center} 
\vskip .4in 
 
\begin{center} {\bf ABSTRACT }  
\end{center} 
\begin{quotation}\noindent  
 We study  N=2 nonlinear two dimensional sigma models with boundaries and their massive generalizations
 (the Landau-Ginzburg models). These models are defined over either K\"ahler or bihermitian 
 target space manifolds. We determine the most general local N=2 superconformal boundary 
conditions (D-branes) for
 these sigma models. In the K\"ahler case we reproduce the known results in a systematic 
 fashion including interesting results concerning the coisotropic A-type branes. 
 We further analyse  the N=2 superconformal boundary conditions
 for sigma models defined over a bihermitian manifold with torsion. 
 We interpret the boundary conditions in terms of different types of submanifolds of the target
 space. We point out how the open sigma models correspond to
 new types of target space geometry.
For the massive Landau-Ginzburg
 models (both K\"ahler and bihermitian) we discuss an important class of supersymmetric
 boundary conditions which admits a nice geometrical interpretation.
\end{quotation} 
\vfill 
\eject 

\tableofcontents
 
\Section{Introduction}

The superconformal boundary conditions that arise from the two-dimensional non-linear sigma model
 are relevant for the description of D-brane in a semiclassical approximation.
 Thus the superconformal boundary conditions of sigma model are of manifest interest for 
  string theory. 

 In the present paper we undertake a study of local N=2 superconformal 
 boundary conditions for the general (2,2) non-linear sigma model. There are two different types of (2,2)
 non-liner sigma model depending on the presence or not of a torsion $H=dB$. In the torsion
 free case the (2,2) sigma model has to be defined over a K\"ahler manifold \cite{Zumino:1979et}.
  The requirement of conformal invariance (i.e., Ricci flatness) restricts  these manifolds
 to be Calabi-Yau manifolds which play a prominent role in string 
 compactification. Starting with the work of \cite{Ooguri:1996ck},  N=2 superconformal 
 boundary conditions for K\"ahler sigma models have been extensively discussed
 in the literature. However it is instructive to return to the problem and rederive the 
 known results in a systematic fashion. Indeed we find  new properties of the 
 N=2 superconformal boundary conditions for the K\"ahler sigma model.
 We believe that our treatment of the K\"ahler sigma model
  is exhaustive with our framework and given our general assumptions.

In the torsionful case the (2,2) sigma model has to be defined over a bihermitian manifold with special 
 properties \cite{Gates:1984nk}. This type of manifold can be thought of as a special kind of 
 modified Calabi-Yau manifold \cite{Rocek:1991ze} and
  may play an important role for string compactifications
 in the presence of non-zero NS-NS three form field stregth (e.g., the N=2 WZW models). 
 We do not know of any previous study 
 of the N=2 superconformal boundary conditions for the sigma model over a bihermitian manifold and 
  present a detailed analysis of this case. 

 There are massive generalizations of N=2 non-linear sigma models (the Landau-Ginzburg models)
 which are (manifestly) not conformally invariant. However they admit a wide class of supersymmetric
  boundary conditions which are similar to the D-brane conditions for the conformal sigma model.
 Hence it is appropriate to discuss the Landau-Ginzburg models in the present context
 (For N=1 boundary conditions are discussed in \cite{rocek}).
 By analysing N=2 Landau-Ginzburg models we clarify certain points concerning
  previous results \cite{Hori:2000ck} and also to extend the analysis to the bihermitian 
 Landau-Ginzburg models. 

 Before going into the detailed discussion of the problem let us briefly examine the notion
 of a boundary condition which respect some symmetries (e.g., supersymmetry). This is a much
  more subtle question than usually appreciated.

 We identify three different routes for finding boundary conditions: By

{\bf *} requiring the boundary conditions to be consistient with the symmetry 
 algebra, e.g. for (on-shell/off-shell) supersymmetry
$$ F(X,\psi)=0\,\,\,\,\,\,\,\rightarrow\,\,\,\,\,\,\,\delta_{susy}F(X,\psi)=0$$

{\bf *} imposing appropriate boundary conditions on conserved currents
 in order to have the conserved charges in the presence of boundaries, e.g.
 for N=1 superconformal symmetry
$$ T_{++}-T_{--}=0\,\,\,\,\,\,\,\,\,\,\,\,\,\,\,\,G_+-\eta G_- =0$$

{\bf *} requiring that the boundary terms of the appropriate variations of the action
 are zero (e.g., the general field variation and the supersymmetric variation)
$$ \delta S = 0,\,\,\,\,\,\,\,\,\,\,\,\,\,\,\,\,\,\delta_{susy}S=0$$

It is important to realize that there are clear distictions between these three
 alternatives. The first algebraic requirement is extremely useful since it relies 
 entirely on the representation theory of a given symmetry. For instance, as we will see
  in the case of supersymmetry there are properties of the boundary conditions which
 are found using only off-shell supersymmetry. Therefore these properties do not depend on
 the details of the model.  However one cannot hope to get all dynamics out of the algebra
 and thus in this respect the algebraic requirements are not complete.

On the other hand the conditions involving currents are dynamical requirements which depend
 only on the bulk properties of the theory. The currents are defined only on-shell
 and thus the boundary conditions for the currents make sense only on-shell.     

 From a technical point of view the action is, perhaps, the simplest way
 to derive  boundary conditions.  
 But using an action is difficult in a generic situation since we have to more or less guess which
  boundary terms to include.
  An example
 of this difficulty is the  N=1 sigma model with non-zero B-field \cite{Haggi-Mani:2000uc}.
 Starting from the standard $N=1$ superfield action one would not find
  boundary conditions which respect supersymmetry on the boundary. 
 One would then have to add boundary B-field terms by hand. Once one has found
  boundary conditions using one of the other approaches, however, one may add boundary terms to the 
 action so as to reproduce them.
Because of these issues
 there are contradictory statements about boundary conditions for the N=1 sigma model 
 with $B$-field in the literature
 (see, for example, \cite{BorlafLozano}).

 Consequently these approaches have their own advantages and disadvantages.
 Nevertheless these three approaches (or combinations of them) 
  may produce a reliable answer. In this article we adopt the following logic.
 We always start from the analysis of the algebraic requirements on the general ansatz for
 boundary conditions on the physical fermions and subsequently study the boundary conditions for  
  currents. As a final step in the analysis we find an action from which the boundary 
 conditions can be derived. 
 To avoid  confusion
 we  discriminate between the \emph{bulk action} 
 (without boundary terms) and the \emph{boundary action} (with boundary terms).
 For the bulk action we are only 
 ineterested in properties up to boundary terms. Thus the bulk action is used to find
  the equations of motion, on-shell supersymmetry 
 transformations and conserved currents. For the derivation 
 of the boundary conditions from an action we use the boundary action. 
  
The article is organised as follows. In Section~\ref{s:review} we review the boundary conditions
 which arise from the requirement of N=1 superconformal invariance. We sketch the main results
 from \cite{Albertsson:2001dv} and \cite{Albertsson:2002qc}. Following this introductory section,
 in Section~\ref{s:Kahlersigma} we treat the
  N=2 sigma model on a K\"ahler manifold. This model has been extensively discussed
 in the literature because of its relevance to the description on D-branes on Calabi-Yau manifolds.
 We rederive the known results in a systematic fashion and find  new
 results. We emphasize the differences between the present analysis and the previous studies
 of this type of boundary conditions. One of the interesting results we discuss is the full description 
 of A-type branes with a $B$-field, as coisotropic submanifolds with 
 $f$-structures on them. 
 In the following Section~\ref{s:bihsigma} we study the N=2 superconformal boundary conditions 
 for the bihermitian sigma model. From a technical point of view this problem is harder 
 than the K\"ahler case partially due to the absence of a special system of coordinates such as
 the canonical complex coordinates in the K\"ahler case. However we derive the general results and give
 a partial  interpretation of them in geometrical terms.
 It seems that a geometrical interpretation of these formal results requires 
 a better mathematical insight into the problem than we have at present. 
 Section~\ref{s:LGmodel} deals with the massive generalization of the models discussed
 in the previous sections. We are mainly concerned with the behaviour of the prepotential 
 for certain supersymmetric solutions that admit a nice geometrical interpretation. 
 Finally, in Section~\ref{s:discussion} we give a summary of the paper with a discussion
 of the open problems and the relation of our result to previous results in the literature. 
There are rather detailed Appendixes on N=1 and N=2 supersymmetry which establish
  our conventions and also point out some subtleties in the N=2 formalism in 
 the presence of boundaries. In three last Appendices some geometrical background is
 presented.

\Section{N=1 review} 
\label{s:review}

In this section we review results on N=1 superconformal boundary conditions 
 for non-linear sigma models. This will allow  us to introduce the notation
 and some relevant concepts. We closely follow ref's \cite{Albertsson:2001dv}
 and \cite{Albertsson:2002qc}.

 The starting point is a general $N=1$ sigma model given by the bulk action
\beq
 S= \int d^2\sigma\,d^2\theta\,\,D_+\Phi^\mu D_- \Phi^\nu (g_{\mu\nu}(\Phi) 
 + B_{\mu\nu}(\Phi))
\equiv\int d^2\sigma\,d^2\theta\,\,D_+\Phi^\mu D_- \Phi^\nu E_{\mu\nu}(\Phi)
\eeq{bulkaction1}
 where  $E_{\mu\nu}\equiv g_{\mu\nu} +B_{\mu\nu}$
 and  we use a N=1 superfield formalism (see Appendix A for conventions).
The model is defined over 
 a Riemannian manifold of dimension $d$ with
$g_{\mu\nu}$ being the Riemannian metric\footnote{Hereafter,by
 ``Riemannian'', we shall mean ``Riemannian or pseudo Riemannian''.} 
 and $B_{\mu\nu}$, a general antisymmetric field.
  The theory is classically conformally invariant and thus there is no dimensionful
 parameter in the classical 
 theory. With the assignment  $d[\sigma]=1$ and $d[\d]=-1$, the canonical dimensions of
  the fields are $d[X]=0$, $d[\psi]=1/2$. A  general ansatz\footnote{We need to assume
 differentiability to be able to do calculations. Thus within this classical 
 approach we have to exclude such brane configurations where this property is 
 lost (e.g., a brane ending on a
 brane).} for the fermionic
 boundary conditions has the following form
\beq
 \psi_-^\mu = {\cal R}^\mu(\psi_+, X) = R^\mu_{\,\,\nu}(X)\psi_+^\nu + 
 R^\mu_{\,\,\nu\rho\sigma}(X)\psi_+^\nu
 \psi_+^\rho \psi_+^\sigma + ... .
\eeq{genansatz}
 However, from  dimensional analysis it may be seen that in the classical theory 
 there are no higher fermion terms. Even for the most general ansatz where 
we allow
 ${\cal R}$ to depend on derivatives of the fields the only possible {\it local}
  dimension $-1/2$ term is the first term on the 
 right hand side of (\ref{genansatz}). Other terms which are allowed by dimensional 
 analysis have the form $\psi^k (\d)^{-1/2(k-1)}$ and are thus  nonlocal. 
 In the quantum theory 
 a mass scale is typically generated and thus higher fermion terms are allowed.
 It is easy to see that the problem may then be solved order by order 
 in the fermions.  

 Since we are interested in the classical boundary conditions we start from the
 following fermionic boundary conditions
\beq
 \psi_-^\mu = \eta R^\mu_{\,\,\nu}(X)\psi^\nu_+ .
\eeq{standanR}
 which are the {\it most general local} fermionic boundary conditions in the classical theory.
For the properties of $R$ we introduce a terminology  which generalizes the well-known concepts 
for D-branes in flat space-time.
  At a given point $X$, by going to special system of coordinates  $R$ can 
 always be  brought to the form 
\beq
 R = \left ( \begin{array}{ll}
 R^m_{\,\,n} & 0 \\
 0 & -\delta^i_{\,\,j}
\end{array} \right ).
\eeq{canonform}
 Thus we may introduce a projector $Q$ at this point ($Q^2=Q$)
\beq
  Q = \left ( \begin{array}{ll}
  0 & 0 \\
 0 & \delta^i_{\,\,j}
\end{array} \right ).
\eeq{Qdef}
 It follows that  $R$, must satsify,  in covariant language,
\beq
 RQ=QR=-Q,\,\,\,\,\,\,\,\,\,\,\,\,\,\,\,\,\,\,\,\,Q^2=Q.
\eeq{propQR}
  A maximal projector $Q$ with the property (\ref{propQR})
 will be called a Dirichlet projector, (justified below).
  If the rank of such a $Q$ is $d-(p+1)$ we say that 
 $R$ corresponds to a Dp-brane. The projector which is complementary 
 to $Q$ is $\pi=I-Q$. We refer to $\pi$ as a Neumann projector. The following 
 properties of $Q$, $\pi$ and $R$ will be useful
\beq
 \pi^2=\pi,\,\,\,\,\,\,\,\,\,\,
 Q^2 =Q,\,\,\,\,\,\,\,\,\,\,
r^2=I,\,\,\,\,\,\,\,\,\,\,
 P\pi =\pi P =P,\,\,\,\,\,\,\,\,\,\,
 PQ=QP=0
\eeq{projectorprop}
 where $r\equiv \pi-Q$ and $2P\equiv I+R$.
 In the absence of a $B$-field\footnote{By $B$-field we shall mean the sum of
 the actual NS $B$-filed and the $U(1)$ field strength.}, the relation $R^2=1$ holds
  and this case corresponds to parity invariant 
 boundary conditions. 

 We consider boundary conditions from the most general point of view,
 the off-shell supersymmetry transformations of the scalar mupliplet. 
Applying the off-shell supersymmetry transformations (\ref{compsusytrN1}) to the ansatz (\ref{standanR})
 we obtain
\beq
 \d_= X^\mu - R^\mu_{\,\,\nu}\d_\+ X^\nu - 2i \eta P^\mu_{\,\,\nu} F^\nu_{+-}
 + 2iP^\sigma_{\,\,\rho} R^\mu_{\,\,\nu,\sigma} \psi_+^\rho \psi_+^\nu = 0 
\eeq{susyoffbos}
 where $\epsilon^-=\eta\epsilon^+$ with $\eta^2=1$ in (\ref{compsusytrN1}).
 Contracting with $Q$ we get 
\beq
 Q^\mu_{\,\,\nu} \d_0 X^\nu + 2i P^\rho_{\,\,\gamma} P^\nu_{\,\,\sigma}
 Q^\mu_{\,\,\nu,\rho} \psi_+^\sigma \psi_+^\gamma = 0 .
\eeq{Qbosbc11}
 To understand this relation better let us jump ahead a bit and discuss the 
 special case when
$Q^\mu_{\,\,\nu}\d_0 X^\nu=0$ (or equivalently $\pi^\mu_{\,\,\nu}\d_0 X^\nu =
 \d_0 X^\mu$). Thus in (\ref{Qbosbc11}) the two-fermion term should 
 vanish by itself implying that
\beq
  P^\mu_{\,\,\nu} P^\rho_{\,\,\sigma} Q^\lambda_{\,\,[\mu,\rho]}=0.
\eeq{inegrabl}
 Because $\pi P= P\pi =P$ we can rewrite the condition 
 (\ref{inegrabl}) in a completly equivalent form as follows
\beq
 \pi^\mu_{\,\,\nu} \pi^\rho_{\,\,\sigma} Q^\lambda_{\,\,[\mu,\rho]}=0
\eeq{inegrabl11}
 which is the integrability condition for $\pi$. 
 Due to this condition the Dp-brane can be interpreted as a maximal integral 
 submanifold and $\d_0 X^\mu$ as living in the tangent space of this submanifold, 
 as was pointed out in \cite{Albertsson:2001dv}. 

It is important
 to realize that the above argument is independent of the details of the model (e.g.,
 on-shell realization of supersymmetry transformation), as it is based only on
 the form of the ansatz (\ref{standanR}) and the off-shell supersymmetry
 transformations for the scalar multiplet. Therefore this geometrical 
 interpretation is still valid, e.g., for the massive sigma model (the Landau-Ginzburg model).
 However for the Landau-Ginzburg model the argument for the uniqueness of the 
 ansatz (\ref{standanR}) breaks down due to the presence of a dimensionful parameter.

 Returning to the N=1 conformal sigma model (\ref{bulkaction1}), the on-shell supersymmetry 
 transformation together with ansatz (\ref{standanR}) fixes the boundary conditions to be of 
 the following form
\beq
\left \{ \begin{array}{l}
 \psi_-^\mu = \eta R^\mu_{\,\,\nu}\psi_+^\nu \\
 \d_= X^\mu - R^\mu_{\,\,\nu} \d_\+ X^\nu + 2i(P^\sigma_{\,\,\gamma} \nabla_\sigma R^\mu_{\,\,\nu}
 + P^\mu_{\,\,\rho} g^{\rho\delta} H_{\delta\sigma\gamma} R^\sigma_{\,\,\nu})\psi_+^\gamma
 \psi_+^\nu =0 .
\end{array}\right .
\eeq{formbconshN1}

 Now we sketch the main steps in the derivation of the general N=1
 superconformal boundary conditions \cite{Albertsson:2002qc}. These will be relevant
 to what follows in that the requirement
 of N=1 superconformal invariance is part of the N=2 superconformal invariance. 
 The components of the supercurrent and the stress tensor that we need may be taken to have
 the following expressions
\beq 
G_{+} = \psi_{+}^\mu \d_{\+} X^\nu g_{\mu\nu} - \frac{i}{3} 
\psi_{+}^\mu \psi_{+}^\nu \psi_{+}^\rho H_{\mu\nu\rho} , 
\eeq{comp1} 
\beq 
G_{-} =  \psi_{-}^\mu \d_{=} X^\nu g_{\mu\nu} + \frac{i}{3} 
\psi_{-}^\mu \psi_{-}^\nu \psi_{-}^\rho H_{\mu\nu\rho} , 
\eeq{comp2} 
\beq 
T_{++} = \d_{\+}X^\mu \d_{\+}X^\nu g_{\mu\nu} + i 
\psi^\mu_+ \nabla^{(+)}_{+} \psi^\nu_{+} g_{\mu\nu} , 
\eeq{comp3} 
\beq 
T_{--} =   \d_{=}X^\mu \d_{=}X^\nu g_{\mu\nu} + 
 i \psi^\mu_- \nabla^{(-)}_{-} \psi^\nu_{-} g_{\mu\nu} , 
\eeq{comp4} 
where the covariant derivatives acting on 
the worldsheet fermions are defined by 
\beq 
\nabla^{(+)}_{\pm}\psi_{+}^\nu = \partial_{\pp}\psi_{+}^\nu + 
\Gamma^{+\nu}_{\,\,\rho\sigma}\d_{\pp} X^\rho 
\psi_{+}^\sigma,\,\,\,\,\,\,\,\,\,\, \nabla^{(-)}_{\pm}\psi_{-}^\nu = 
\partial_{\pp}\psi_{-}^\nu + \Gamma^{-\nu}_{\,\,\rho\sigma}\d_{\pp} 
X^\rho \psi_{-}^\sigma . 
\eeq{covdervferm}
 To ensure N=1 superconformal invariance we have to impose the 
 following conditions on the above currents on the boundary
\beq
 T_{++}-T_{--}=0,\,\,\,\,\,\,\,\,\,\,\,\,\,\,\,\,\,\,\,
 G_{+}-\eta G_{-}=0 .
\eeq{boundcurN2}
 Classically these conditions make sense only on-shell since the conserved currents 
 are defined modulo the equations of motion. 
 Thus we should make use
 of the field equations in our analysis. Using the fermionic equations of motion,
\beq 
g_{\mu\nu} (\psi_{+}^\mu \nabla^{(+)}_- \psi_+^\nu - \psi_{-}^\mu \nabla^{(-)}_+ 
\psi_-^\nu) = 0, 
\eeq{fermeqdbo}
 we substitute the fermionic ansatz in the conformal condition and get
\ber 
\nonumber 
0=T_{++} - T_{--} &=& 2i \psi_{+}^\sigma\d_0\psi_{+}^\lambda \left[ 
g_{\sigma\lambda} - R^\mu_{\,\,\sigma} g_{\mu\nu} R^\nu_{\,\,\lambda} 
 \right] + \\ 
\nonumber & +& 2\d_0 X^\delta \pi^\rho_{\,\,\delta} \left[ g_{\rho\nu} 
(\d_\+ X^\nu - \d_= X^\nu) - 2 B_{\rho\nu} \pi^\nu_{\,\,\lambda} \d_0 
X^\lambda  +\right. \\ 
\nonumber  &+& \left. i \left( R^\mu_{\,\,\gamma}\Gamma_{\mu\rho\nu} 
 R^\nu_{\,\,\sigma} -\Gamma_{\gamma\rho\sigma} 
 - R^\mu_{\,\,\sigma} g_{\mu\nu}  
R^\nu_{\,\,\gamma,\rho} + \right. \right. \\ 
&+& \left. \left. H_{\sigma\rho\gamma} + R^\mu_{\,\,\sigma} 
 H_{\mu\rho\nu} R^\nu_{\,\,\gamma} \right) \, \psi^\sigma_+ \psi^\gamma_+  
   \right] , 
\eer{TTTterms} 
 where we assume that the string is confined to some of the  directions (i.e., that there 
 is a projector $Q$, $Q^2=Q$ such that $Q^\mu_{\,\,\nu}\d_0 X^\nu=0$
 or $\pi^\mu_{\,\,\nu}\d_0 X^\nu =\d_0 X^\mu$, $\pi\equiv I-Q$). In (\ref{TTTterms})
 we introduce an antisymmetric tensor $B_{\mu\nu}$ which a priori has nothing 
 to do with $H$. The term which contains $B$ vanishes identically in this expression and represents
 the ambiguity in the right hand side.  We further stress
 that we do not assume any relation between $Q$ and $R$ (a condition
 $RQ=QR=-Q$ will arise from the supersymmetry requirements below). From the condition
 (\ref{TTTterms}) we get the conformal boundary
 conditions
\beq 
\left \{ 
\begin{array}{l} 
 \psi_-^\mu - \eta R^\mu_{\,\,\nu}\psi_+^\nu = 0, \\
 \pi^\rho_{\,\,\delta} E_{\nu\rho} \pi^\nu_{\,\,\lambda} 
 \d_\+ X^\lambda -  \pi^\rho_{\,\,\delta} E_{\rho\nu}  
\pi^\nu_{\,\,\lambda} \d_= X^\lambda 
- i\pi^\rho_{\,\,\delta} (R^\mu_{\,\,\sigma} g_{\mu\nu} \nabla_\rho 
  R^\nu_{\,\,\gamma}  -  H_{\sigma\rho\gamma} - R^\mu_{\,\,\sigma} 
 H_{\mu\rho\nu} R^\nu_{\,\,\gamma} )\psi^\sigma_+ \psi^\gamma_+ = 0, \\ 
 Q^\mu_{\,\,\lambda}(\d_= X^\lambda  + \d_\+ X^\lambda)= 0, 
\end{array} \right . 
\eeq{bosb111} 
 together with the condition
\beq
R^\mu_{\,\,\sigma} g_{\mu\nu} R^\nu_{\,\,\rho} = 
g_{\sigma\rho} .
\eeq{RgRgcond}
 Substituting these relations into the second condition in (\ref{boundcurN2}),
 $G_+ -\eta G_- =0$, we find the 
  relations
\beq 
\left \{ 
\begin{array}{l} 
Q^\mu_{\,\,\nu} R^\nu_{\,\,\rho} = R^\mu_{\,\,\nu} Q^\nu_{\,\,\rho} = - Q^\mu_{\,\,\rho} \\
 \pi^\rho_{\,\,\delta} E_{\nu\rho} \pi^\nu_{\,\,\gamma} = \pi^\rho_{\,\,\delta} 
 E_{\rho\nu} \pi^\nu_{\,\,\lambda} R^\lambda_{\,\,\gamma} \\
\pi^\mu_{\,\,\gamma} 
\pi^\rho_{\,\,\phi} \nabla_{[\rho} Q^\delta_{\,\,\mu]} = 0 \\ 
\pi^\mu_{\,\,\tau} \pi^\nu_{\,\,\sigma} \pi^\rho_{\,\,\gamma} 
H_{\mu\nu\rho} = \frac{1}{2} \pi^\mu_{\,\,\tau} \pi^\nu_{\,\,\sigma} 
\pi^\rho_{\,\,\gamma} (\nabla_\mu B^{\cal D}_{\nu\rho} + \nabla_\nu 
B^{\cal D}_{\rho\mu} 
+ \nabla_\rho B^{\cal D}_{\mu\nu}) 
\end{array} \right . 
\eeq{propN1susy} 
 where $B^{\cal D}\equiv \pi^t B\pi + Q^t B Q$. (Alternatively in the last line
 of (\ref{propN1susy}) $B^{\cal{D}}$ 
 may be replaced by $B^{\pi}\equiv \pi^t B\pi$.) If  $H=dB$ then  the last property
 in (\ref{propN1susy}) would be trivially satisfied because of the integrability of
 $\pi$ (the next to last condition). Using the properties (\ref{RgRgcond}) and (\ref{propN1susy}) 
 we see that the 
 result is compatible with the supersymmetry algebra (i.e., we may rewrite 
 the bosonic conditions in (\ref{bosb111}) in the form they have in (\ref{formbconshN1})).
In what follows we shall sometimes need
\beq
 R^\mu_{\,\,\nu} = r^\mu_{\,\,\nu} - 2g^{\mu\lambda} B^{\pi}_{\lambda\rho} P^\rho_{\,\,\nu} ,
\eeq{PrRform}
 which follows from (\ref{propN1susy}).

All the above results can be derived differently, starting from an appropriate action
 which we now describe.

 If we look for parity invariant boundary conditions (i.e., $R^2=I$) we must start 
 from an action without explicit boundary terms
\beq
 S=  \int d^2\xi d^2\theta\,\, D_{+} \Phi^\mu D_{-} \Phi^\nu g_{\mu\nu} 
(\Phi) .
\eeq{boundRIactN1}
 Requiring that the boundary field equations are satified as well as invariance under supersymmetry, 
including the boundary, we
 reproduce the N=1 superconformal boundary conditions.

 In the general case with non-zero $B$ we have to use
 the following action with explicit boundary terms,
\beq
 S= \int d^2\xi d^2\theta\,\, D_{+} \Phi^\mu D_{-} \Phi^\nu E_{\mu\nu} 
(\Phi) -\frac{i}{2}\int d^2\xi\,\,  \partial_{=}( B_{\mu\nu}\psi_{+}^\mu 
\psi_{+}^\nu + B_{\mu\nu} \psi_-^\mu \psi_-^\nu) . 
\eeq{boundBN1act}
 In contrast to the current analysis we do not have to assume that 
 $H=dB$ here since this is so by construction.
 For details of the derivation see 
 \cite{Albertsson:2001dv} and \cite{Albertsson:2002qc}.

The literature does not agree on the presence of the two-fermion terms in 
 the bosonic boundary conditions. Therefore we would like to elaborate on this point.   
 In the previous discussion we have outlined the proof that there is a one to one 
 correspondence between the classical N=1 superconformal boundary conditions of 
 the sigma model and submanifolds with a $B$-field (the $\pi$-integrable $r$ and the $B$-term 
 in (\ref{PrRform})). If we relax the requirements
 of N=1 superconformal invariance to conformal invariance  only, or to supersymmetric 
 invariance only, the correspondence would no-longer be one to one. For example, if we adopt the boundary conditions
 $\psi_-=\eta R\psi_+$ and $\d_= X = R \d_\+X$, then   
 the supersymmetry current conditions in (\ref{boundcurN2}) are satisfied (with $H=0$)
\beq
 G_{+} -\eta G_{-} = \psi_+^\mu g_{\mu\nu} \d_\+ X^\nu -\eta \psi_-^\mu g_{\mu\nu} \d_= X^\nu 
 = 0
\eeq{susycurexam}
 provided that $R^t g R=g$. To interpret these boundary conditions as representing submanifolds
 of the target manifold, however, we have to add the inegrability conditions by hand. Further, unless
 extra requirement are imposed on $R$ these solutions are not conformally invariant
 (i.e., $T_{++}-T_{--}\neq 0$). The extra requirements derived in this manner are much stronger than
 those previously discussed. Hence we see how different 
 requiremens on the boundary conditions lead to the different results and
  interpretations. This partially explains the disagreement in the literature.

However if we look at {\it the N=1 superconformal boundary conditions} (\ref{formbconshN1}) 
 (together with the properties (\ref{RgRgcond}) and (\ref{propN1susy})) 
 the two fermion terms \emph{are} there  
 in the general solution. 
 It still makes sense, of course to ask under which circumstances the two-fermion term is absent
 in the boundary conditions. There is a simple theorem, for the parity 
 invariant boundary conditions, which shows
 that the two-fermion term is absent in the bosonic boundary condition if and only if
 the submanifold is totally geodesic 
 (for the details we refer to the Appendix C).
 If the two-fermion term vanishes then 
 the following property holds on the boundary
\beq
 \delta X^\mu g_{\mu\nu} \d_1 X^\nu =0 .
\eeq{deltaXXbound} 
 Thus we see that the property (\ref{deltaXXbound}) for the parity invariant N=1 superconformal 
 boundary conditions is equivalent to the statement that the corresponding submanifold is totally 
 geodesic. In fact the property of being totally geodesic has a very simple physical 
 interpretation; once a geodesic starts along the submanifold   it cannot escape the submanifold
 since the second fundamental form is identically zero. Therefore, classically, particles cannot 
 escape from the D-brane in this case.

\Section{N=2 sigma models on K\"ahler manifolds} 
\label{s:Kahlersigma}

 We start our analysis from the relatively well studied example of an N=2 sigma model 
 with action
\beq
 S= \int d^2\sigma\,d^2\theta\,\,D_+\Phi^\mu D_- \Phi^\nu g_{\mu\nu}(\Phi).
\eeq{action}
 This action is written using N=1 superfield notation and is thus manifestly N=1
 supersymmetric. For irreducible even dimensional manifolds the action (\ref{action})
  has an additional (nonmanifest) 
 supersymmetry if and only if the manifold is K\"ahler \cite{Zumino:1979et}. The extra
 supersymmetry transformation is
\beq
 \delta_2 \Phi^\mu = \epsilon^\alpha_2 D_\alpha \Phi^\nu \,J^\mu_{\,\,\nu}(\Phi)
\eeq{extrasusyK}
 where $J^\mu_{\,\,\nu}$ is the covariantly constant complex
 structure of the manifold. The metric $g_{\mu\nu}$ should be Hermitian with respect to
 this complex structure. The action (\ref{action}) can be rewritten in a manifestly N=2 supersymmetric
 form as
\beq
 S= \int d^2\sigma\,d^2\theta\,d^2\bar{\theta}\,\,K(\Phi, \bar{\Phi})
\eeq{Kaction}
 where the superfields are now complex and $K$ is a K\"ahler potential for the Hermitian metric. 

The boundary conditions for this model were first considered in \cite{Ooguri:1996ck}. Here
 we reproduce some of their results from a different point of view and  
 complement them by new ones.

\subsection{The N=2 algebra}
\label{s:N2algebra}
 
A lot of information about the N=2 supersymmetric boundary conditions can be deduced by
 purely algebraic considerations. In components the manifest on-shell\footnote{The auxiliary
 (F-)field is integrated out.} supersymmetry transformations
 are
\beq
\left \{ \begin{array}{l}
 \delta_1 X^\mu = - (\epsilon^+_1 \psi_+^\mu + \epsilon_1^- \psi_-^\mu) \\
 \delta_1 \psi_+^\mu = -i \epsilon_1^+ \d_\+ X^\mu + \epsilon_1^- 
 \Gamma^{\mu}_{\,\,\nu\rho} \psi_-^\rho \psi_+^\nu \\
 \delta_1 \psi_-^\mu = -i\epsilon_1^- \d_= X^\mu - \epsilon_1^+ \Gamma^{\mu}_{\,\,\nu\rho}
 \psi_-^\rho \psi_+^\nu
\end{array}
\right .
\eeq{n11}
 and the nonmanifest tranformations (\ref{extrasusyK}) are
\beq
\left \{\begin{array}{l}
 \delta_2 X^\mu =  (\epsilon^+_2 \psi_+^\nu + \epsilon_2^- \psi_-^\nu) J^\mu_{\,\,\nu} \\
 \delta_2 \psi_+^\mu = -i \epsilon_2^+ \d_\+ X^\nu J^\mu_{\,\,\nu} - \epsilon_2^- 
 J^\mu_{\,\,\sigma}\Gamma^{\sigma}_{\,\,\nu\rho} \psi_-^\rho \psi_+^\nu 
 - \epsilon_2^+ J^\mu_{\,\,\nu,\rho} \psi_+^\rho \psi_+^\nu - \epsilon_2^- J^\mu_{\,\,\nu,\rho}
 \psi_+^\rho \psi_-^\nu  \\
 \delta_2 \psi_-^\mu = -i\epsilon_2^- \d_= X^\nu J^\mu_{\,\,\nu} 
 +\epsilon_2^+ J^\mu_{\,\,\sigma}\Gamma^{\sigma}_{\,\,\nu\rho}
 \psi_-^\rho \psi_+^\nu -\epsilon_2^+ J^\mu_{\,\,\nu,\rho} \psi_-^\rho \psi_+^\nu
 - \epsilon_2^- J^\mu_{\,\,\nu,\rho}\psi_-^\rho \psi_-^\nu
\end{array}
\right .
\eeq{n12}
A supersymmetry transformation of the  fermionic boundary conditions
\beq
\psi_-^\mu = \eta_1 R^\mu_{\,\,\nu}(X) \psi_+^\nu
\eeq{ferman}
yields
\beq
 \delta_i \psi_-^\mu = \eta_1 R^\mu_{\,\,\nu,\rho}\delta_i X^\rho \psi_+^\nu +
 \eta_1 R^\mu_{\,\,\nu}\delta_i\psi_+^\nu,\,\,\,\,\,\,\,\,\,\,\,\,i=1,2 .
\eeq{variat}
 This is the corresponding bosonic boundary conditions. 
 
 The first supersymmetry variation (\ref{n11}) applied to (\ref{ferman}) yields
 the expression
 discussed in the previous section
\beq
 \d_= X^\mu - R^\mu_{\,\,\nu} \d_\+ X^\nu   + 2i   
 P^\rho_{\,\,\gamma} \nabla_\rho R^\mu_{\,\,\nu}  \psi_+^\gamma \psi_+^\nu =0 
\eeq{stanres}
 where $\epsilon_1^- = \eta_1 \epsilon_1^+$ ($\eta_1^2=1$) and $2P^\mu_{\,\,\nu}=\delta^\mu_{\,\,\nu}
 + R^\mu_{\,\,\nu}$.
 The second 
 supersymmetry variation  (\ref{n12}) yields
\beq
 \d_= X^\mu + (\eta_1\eta_2)J^\mu_{\,\,\sigma} R^\sigma_{\,\,\nu} J^\nu_{\,\,\gamma} 
 \d_\+ X^\gamma + i\left [ (\eta_1\eta_2) J^\mu_{\,\,\sigma}
 \nabla_\rho R^\sigma_{\,\,\nu} J^\rho_{\,\,\gamma} +
   J^\mu_{\,\,\sigma} \nabla_\rho R^\sigma_{\,\,\nu} J^\rho_{\,\,\lambda} 
 R^\lambda_{\,\,\gamma}  \right ] \psi_+^\gamma \psi_+^\nu =0
\eeq{susy2aa}
 where we used that $\epsilon_2^+=\eta_2\epsilon_2^-$ and
  the fact that $\nabla_\rho J^\mu_{\,\,\nu}=0$.

 Equations (\ref{stanres}) and (\ref{susy2aa}) should be equivalent.
 Comparing the $X$-part we get the following condition 
\beq
 J^\mu_{\,\,\gamma} R^\gamma_{\,\,\nu} J^\nu_{\,\,\sigma} = - (\eta_1\eta_2) R^\mu_{\,\,\sigma}
\eeq{compbos} 
 or equivalently 
\beq
 J^\mu_{\,\,\gamma} R^\gamma_{\,\,\nu} = (\eta_1\eta_2) R^\mu_{\,\,\gamma} J^\gamma_{\,\,\nu}.
\eeq{compbos2}
 Using (\ref{RgRgcond}), i.e. that $R^\mu_{\,\,\rho} g_{\mu\nu} R^\nu_{\,\,\sigma} = g_{\rho\sigma}$,
 this equation can be equivalently rewritten as follows
\beq
 R^\mu_{\,\,\rho} J_{\mu\nu} R^\nu_{\,\,\sigma} = (\eta_1\eta_2) J_{\rho\sigma} .
\eeq{kahformRR}
 The case $\eta_1\eta_2=1$ corresponds to the so called B-type conditions and $\eta_1\eta_2=-1$
 to the A-type conditions, as defined in \cite{Ooguri:1996ck}. 
   
 Using  (\ref{compbos}) we rewrite the equation (\ref{susy2aa}) as follows
\beq
 \d_= X^\mu - R^\mu_{\,\,\nu}  
 \d_\+ X^\nu + 2i (\eta_1\eta_2)  J^\mu_{\,\,\lambda} J^\rho_{\,\,\gamma} 
  P^\sigma_{\,\,\rho}  \nabla_\sigma R^\lambda_{\,\,\nu}  
  \psi_+^\gamma \psi_+^\nu =0 .
\eeq{susy2bb}
 Comparing the two-fermion terms in (\ref{stanres}) and (\ref{susy2bb}) we obtain
\beq
 \left ( P^\rho_{\,\,\gamma} \nabla_\rho R^\mu_{\,\,\nu} - (\eta_1\eta_2) 
 J^\mu_{\,\,\lambda} J^\rho_{\,\,\gamma} P^\sigma_{\,\,\rho}  \nabla_\sigma R^\lambda_{\,\,\nu}  
 \right ) \psi_+^\gamma \psi_+^\nu =0 .
\eeq{condtwo}
 Using (\ref{compbos2}) and the antisymmetry of the fermions, (\ref{condtwo}) may
 be rewritten as
\beq
 P^\rho_{\,\,[\gamma|} \nabla _\rho R^\mu_{\,\,|\nu]} - J^\rho_{\,\,[\gamma|}
 P^\sigma_{\,\,\rho} \nabla_\sigma R^\mu_{\,\,\lambda} J^\lambda_{\,\,|\nu]} =0
\eeq{algeintegrablK}
 Finally,  introducing the projectors $\Omega_{\pm} =1/2(I\pm iJ)$ the condition (\ref{algeintegrablK})
 takes the form
\beq
\Omega^\rho_{\pm[\gamma|}
 P^\sigma_{\,\,\rho} \nabla_\sigma R^\mu_{\,\,\lambda} \Omega^\lambda_{\pm|\nu]} = 0.
\eeq{algeintegrablK11}
 This condition does not imply that the two-fermion term vanishes. However it requires
 two-fermion term to have a form which is compatible with the $U(1)$ R-symmetry (see discussion 
 in subsection \ref{sub:currents}).

So far, except in (\ref{kahformRR}), we have used the N=2 supersymmetry algebra only. As the next 
 step we combine the N=2 algebraic results (\ref{compbos})
  and (\ref{algeintegrablK}) with the N=1 current analysis.
 Indeed, as will be shown later, this gives us the full information about
  the N=2 superconformal boundary conditions.  

\subsubsection{B-type}

We first consider the B-type conditions and combine the algebraic requirements 
 (\ref{compbos}) and (\ref{condtwo}) for $\eta_1\eta_2=1$ with the results from the 
 Section~\ref{s:review}.
 On matrix form, the condition (\ref{compbos2}) reads
\beq
 JR=RJ .
\eeq{Btcond}
 Using the Neumann and Dirichlet projectors and their properties (\ref{projectorprop}),
 it follows from (\ref{Btcond})
  that 
\beq
 QJ\pi = \pi J Q=0,\,\,\,\,\,\,\,\,\,\,\,\,\,\,\,
 \pi J = J\pi,\,\,\,\,\,\,\,\,\,\,\,\,\,\,\,
 QJ=JQ.
\eeq{projBtype}
 These expressions are completly equivalent to the statement that $J=rJr$.
 From (\ref{projBtype}) it may be shown that $R$ corresponds
  to a D-brane with odd $p$ (i.e., the world-volume 
 of the brane is even dimensional). From equation (\ref{PrRform}) it follows that
\beq 
 J(\pi^t B\pi)J= (\pi^t B \pi).
\eeq{propBfBtyep}
 The conditions (\ref{condtwo}) are automatically satisfied due to the integrability and
 the property (\ref{RgRgcond}).  

It is useful to rewrite the  boundary conditions  in  the canonical complex  
 coordinates for the complex structure
  such that $J^i_{\,\,j}=i\delta^i_{\,\,j}$, $J^{\bar{i}}_{\,\,\bar{j}}=
 - i\delta^{\bar{i}}_{\,\,\bar{j}}$ and $J^{i}_{\,\,\bar{j}}= J^{\bar{i}}_{\,\,j} =0$.
The condition (\ref{propBfBtyep}) implies that $B^{\pi}_{ij}=B^{\pi}_{\bar{i}\bar{j}}=0$, 
 where the fields are complexified in the standard fashion.
 In these coordinates 
 (\ref{compbos}) implies
\beq
 R^i_{\,\,\bar{j}} = R^{\bar{i}}_{\,\,j}=0,\,\,\,\,\,\,\,\,\,\,\,\,\,
 \nabla_\rho R^i_{\,\,\bar{j}} = \nabla_\rho R^{\bar{i}}_{\,\,j}=0.
\eeq{holR}
 The conditions that follow from (\ref{condtwo}) take the form
\beq
 P^s_{\,\,l} \nabla_s R^i_{\,\,j} = P^s_{\,\,j} \nabla_s R^i_{\,\,l},\,\,\,\,\,\,\,\,\,\,\,\,\,\,\,\,\,
 P^{\bar{s}}_{\,\,\bar{l}} \nabla_{\bar{s}} R^{\bar{i}}_{\,\,\bar{j}} =
 P^{\bar{s}}_{\,\,\bar{j}} \nabla_{\bar{s}} R^{\bar{j}}_{\,\,\bar{l}} .
\eeq{PPQpropK}  
 Using the relations (\ref{PrRform}) and (\ref{propBfBtyep}) it may be seen that (\ref{PPQpropK}) is
 equivalent to the following expressions
\beq
 \pi^s_{\,\,l} \pi^i_{\,\,j} \nabla_s Q^j_{\,\,k}=0,\,\,\,\,\,\,\,\,\,\,\,\,\,\,\,\,\,
\pi^{\bar{s}}_{\,\,\bar{l}} \pi^{\bar{i}}_{\,\,\bar{j}} \nabla_{\bar{s}} 
 Q^{\bar{j}}_{\,\,\bar{k}}=0
\eeq{AformRhol}
 In complex coordinates the integrability
 conditions for $\pi$ (the third property in (\ref{propN1susy})) read
\beq
\begin{array}{ll}
 \pi^j_{\,\,[k} \pi^l_{\,\,p]} \nabla_l Q^i_{\,\,j}=0, &
\pi^j_{\,\,k} \pi^{\bar{l}}_{\,\,\bar{p}} \nabla_{\bar{l}} Q^i_{\,\,j}=0,\\
\pi^{\bar{j}}_{\,\,[\bar{k}} \pi^{\bar{l}}_{\,\,\bar{p}]} \nabla_{\bar{l}} 
 Q^{\bar{i}}_{\,\,\bar{j}}=0, &
\pi^s_{\,\,k} \pi^{\bar{j}}_{\,\,\bar{l}} \nabla_s Q^{\bar{i}}_{\,\,\bar{j}} = 0
\end{array}
\eeq{inegrBincompc}  
 and thus $\pi^i_{\,\,j}$, $\pi^{\bar{i}}_{\bar{j}}$ are  independently integrable.
 Using (\ref{RgRgcond}) we see that (\ref{AformRhol}) follows
 from (\ref{inegrBincompc}). Indeed the condition (\ref{AformRhol}) is weaker than 
 inegrability of $\pi$.

 Summarizing the above, 
  in complex cordinates we have 
 two sets of $d/2$ fermionic and $d/2$ bosonic boundary conditions. One set is
\beq
\left \{ \begin{array}{l}
\psi_-^i = \eta_1 R^i_{\,\,j} \psi_+^j \\
 \d_= X^i - R^i_{\,\,j} \d_\+ X^j  + 4 i P^{\bar{k}}_{\,\,\bar{l}} P^i_{\,\,j} 
 \nabla_{\bar{k}} Q^j_{\,\,s} \psi_+^s \bar{\psi}_+^{\bar{l}}=0  
\end{array} \right .
\eeq{holbc}
 and the other set is obtained by interchanging
 the bared and unbared indices.
 
Geometrical aspects of the above results will be discussed in 
 subsection~\ref{s:geometryK}.

\subsubsection{A-type}

We now turn to the A-type boundary conditions. In this case the condition (\ref{compbos2}) has
 the form
\beq
 JR=-RJ
\eeq{propjsjsjsj}
 We begin with the case without $B$-field (i.e., $R^2=1$). 
Using the properties (\ref{projectorprop}) and (\ref{propjsjsjsj}) it may be seen that 
\beq
 QJQ=0,\,\,\,\,\,\,\,\,\,\,\,\,\,\,\,\,
 \pi J \pi =0\,\,\,\,\,\,\,\,\,\,\,\,\,\,\,\
 rJr = -J.
\eeq{projectAtype}
 From (\ref{projectAtype}) it follows that rank of $Q$ is $d/2$ and so rank$ (\pi) =$
 rank$(Q)$.   
 For this case  the conditions (\ref{condtwo}) is completly equivalent to 
 inegrability of $\pi$: Using the inegrability of $\pi$ we may show that (\ref{condtwo})
 is indeed true. Conversely, contracting (\ref{condtwo}) with $P^\gamma_{\,\,\phi}$ 
 and $P^\nu_{\,\,\epsilon}$ and using the property (\ref{projectAtype}) we recover
  the inegrability of $\pi$ (for this case $\pi=P$). 
 Moreover for this case of so called middle-dimensional branes we cannot introduce
  a $B$-field.

Next we consider the boundary conditions with a $B$-field. Now (\ref{propjsjsjsj}) together 
 with (\ref{projectorprop}) and (\ref{PrRform}) imply 
\beq
 QJQ=0,\,\,\,\,\,\,\,\,\,\,\,\,\,\,
 QJB^\pi =0
\eeq{BIrj1}
 where $B^\pi$ is the field along the brane. In this case at the point we can always 
 bring $R$ to the following block diagonal form
\beq
 R^\mu_{\,\,\nu} = \left (\begin{array}{lll}
  K^\alpha_{\,\,\beta} & 0 & 0 \\
   0 & \delta^n_{\,\,m} & 0 \\
  0 & 0 & -\delta^i_{\,\,j} 
 \end{array} \right )
\eeq{BIrj2}
 where $(ij)$ are the Dirichlet directions, $(\alpha\beta)$ are the Neumann directions
 along which $B^\pi$ is non zero and $(nm)$ are the remaining Neumann directions. 
 In (\ref{BIrj2}) $(I\pm K)^\alpha_{\,\,\beta}$ is invertible in the $(\alpha\beta)$-subspace, 
 i.e. $rank(K)=rank(B^\pi)$.
 Combining (\ref{BIrj1}) and (\ref{BIrj2}) with  the property
\beq
 R^\mu_{\,\,\rho} J_{\mu\nu} R^\nu_{\,\,\lambda} = - J_{\rho\lambda}
\eeq{BIrjr2}
 we find the following form for $J_{\mu\nu}$
\beq
 J_{\mu\nu} = \left (\begin{array}{lll}
  J_{\alpha\beta} & 0 & 0 \\
   0 & 0 & J_{in} \\
  0 & -J_{in} & 0 
 \end{array} \right ).
\eeq{BIrj34}
 Since $J_{\mu\nu}$ is a non-degenerate antisymmetric matrix we must have
\beq
 \frac{1}{2} \left ( \rank(\pi) - \rank(K) + \rank(Q)\right ) = \rank(Q)
\eeq{BIrtr23}
 or alternatively 
\beq
 \rank(\pi)=\frac{1}{2} \left ( d + \rank(B^\pi) \right ).
\eeq{mdBIrj7}
 By definition $\rank(\pi)$ is the dimension of the world-volume of the brane. Thus we find
 that the dimensionality of the A-type D-brane  crucially depends on the rank of the $U(1)$ 
 field strength of the brane. Because  $\rank (\pi^t J\pi) = \rank(B^\pi)$ the pull-back of
 $J_{\mu\nu}$ to the brane is degenerate unless we deal with space-filling brane.
 However there is a non-trivial relation between the geometry of the brane and the allowed 
 $B$-field. Let us look at the details of this relationship for the case of space-filling
 branes ($\rank(B^\pi)=d$). For a space-filling brane the $R$-matrix can be written
 as follows
\beq
 R= \frac{1}{I+\hat{B}} (I-\hat{B})
\eeq{Bfl34}
 where $\hat{B}=g^{-1}B$ and $\hat{B}$ is invertible. Comparing (\ref{Bfl34}) with (\ref{propjsjsjsj})
 we arrive at the relation
\beq
J \hat{B} J\hat{B}  =-I
\eeq{newBI23}
 which in its turn implies that $\tilde{J}^\mu_{\,\,\nu}\equiv 
 J^{\mu\lambda}B_{\lambda\nu}$ is an almost 
 complex stucture. 
 The corresponding Nijenhuis tensor for $\tilde{J}$ is
\beq
 {\cal N}^\mu_{\,\,\nu\rho} (\tilde{J}) = \tilde{J}^\gamma_{\,\,\nu} \nabla_{[\gamma}
 \tilde{J}^\mu_{\,\,\rho]} - \tilde{J}^\gamma_{\,\,\rho} \nabla_{[\gamma}
 \tilde{J}^\mu_{\,\,\nu]} = J^{\mu\lambda} \nabla_\lambda J_{\nu\rho} =0 ,
\eeq{NtennewJJJ}
 where we used that $J$ is covariantly constant and $dB=0$. 
 Thus the new almost complex structure $\tilde{J}$ is integrable.
 Using the definition of $\tilde{J}$ one finds
\beq
 \tilde{J}^\mu_{\,\,\lambda} J_{\mu\rho} \tilde{J}^\rho_{\,\,\sigma} = -J_{\lambda\sigma} ,
\eeq{jjj2222}
 i.e. the K\"ahler form $J_{\mu\nu}$ is a (2,0)+(0,2) form with respect to the new complex 
 structure $\tilde{J}$. Since the K\"ahler form $J_{\mu\nu}$ is nondegenerate we
 need the dimension of ${\cal M}$ to be multiple of 4.
 These are all requirements which follow from the superconformal invariance 
 for the A-type space-filling brane.  
 
 The above structure is realized in some well-known situations. 
 For example, assuming that $\tilde{J}$ is compatible with the metric (i.e., $\tilde{J}^tg\tilde{J} =g$)
 the relation (\ref{jjj2222}) implies that $\{\tilde{J}, J\}=0$. Thus there is one extra 
 almost complex structure $\hat{B}$, ($\hat{B}^2=-I$).
 This situation may be realized for 4k-dimesional manifolds, e.g. for hyperK\"ahler manifolds.
 In this case the three complex structures $(J, \tilde{J}, \hat{B})$ can be shown to
  satisfy the standard $SU(2)$ algebra. 

 In the non extreme cases ($0 < \rank(B^\pi) < d$) the pull back of $J_{\mu\nu}$ is degenerate on the brane.
 However due to the property (\ref{BIrj1}) the complex structures $J$
 restricted to the tangent vectors satisfies the following equation
\beq
 (J^\pi)^3+ J^\pi =0
\eeq{pullfstJ}
 where $J^\pi \equiv \pi J \pi$. Equation (\ref{pullfstJ}) is the generalization of the 
 equation $J^2 = -I$ to the degenerate matrices.  
 Analysing the structure of $R$ in this case we further find that
  $\hat{B}^\pi \equiv \pi g^{-1} B \pi$ satisfies the condition
\beq
 (\hat{B}^\pi J^\pi)^3+ \hat{B}^\pi J^\pi =0
\eeq{pullfstJ11}
 The kernels of $B^\pi_{\mu\nu}$ and of the pull back of $J_{\mu\nu}$ coincide
 by construction. Thus, as in the discussion of (\ref{jjj2222}), we see that 
  $\rank(B^\pi)$ is a multiple of 4.
We  comment more on the geometrical 
 interpretation these  solution in subsection \ref{s:geometryK}. When  $B^\pi$ is
 of maximal rank,  equation (\ref{pullfstJ11}) is equivalent to (\ref{newBI23}).

 As in the B-type case, it is useful to rewrite boundary conditions in the  canonical complex
  coordinates.
The relation
 (\ref{compbos}) now implies
\beq
 R^i_{\,\,j} = R^{\bar{i}}_{\,\,\bar{j}}=0,\,\,\,\,\,\,\,\,\,\,\,\,\,
 \nabla_\rho R^i_{\,\,j} = \nabla_\rho R^{\bar{i}}_{\,\,\bar{j}}=0,
\eeq{holR}
 and  the integrability conditions for $\pi$ become
\beq
\nabla_{[\bar{s}} R^{i}_{\,\,\bar{j}]}=0,\,\,\,\,\,\,\,\,\,
 R^l_{\,\,[\bar{s}|} \nabla_l R^i_{\,\,|\bar{j}]} = 0 ,\,\,\,\,\,\,\,\,\,
 \nabla_{[s} R^{\bar{i}}_{\,\,j]}=0  ,\,\,\,\,\,\,\,\,\,
R^{\bar{l}}_{\,\,[s|}\nabla_{\bar{l}} 
 R^{\bar{i}}_{\,\,|j]}=0,\,\,\,\,\,\,\,\,\,
\eeq{inegAKca}
 
 In complex cordinates the fermionic and bosonic boundary conditions thus read
\beq
\left \{ \begin{array}{l}
\psi_-^i = \eta_1 R^i_{\,\,\bar{j}} \bar{\psi}_+^{\bar{j}} \\
 \d_= X^i - R^i_{\,\,\bar{j}} \d_\+ \bar{X}^{\bar{j}}  + i (\nabla_s R^i_{\,\,\bar{j}} +
 R^{\bar{l}}_{\,\,s}\nabla_{\bar{l}} R^i_{\,\,\bar{j}}) \psi_+^s \bar{\psi}_+^{\bar{j}} = 0
\end{array} \right .
\eeq{holbcAAAA}
 and the other set of  boundary conditions is again obtained by interchanging
 the bared and unbared indices.

Geometrical aspects of these results are  discussed in subsection~\ref{s:geometryK}.

\subsection{Currents}
\label{sub:currents}

 Next we would like to rederive all the previous results in an alternative way: by imposing 
 the boundary conditions on the appropriate currents.

 We want to retain the classical N=2 superconformal invariance in the presence of boundaries.
 Therefore the appropriate objects to study are the currents that correspond to supertranslations
 in (2,2) superspace. However it is instructive to look at the corresponding currents in the 
 N=1 formalism. This has the advantage that we are not confined to the use of complex coordinates.
 The currents are 
\beq
 T^{-}_{\+}=D_+ \Phi^\mu \d_\+ \Phi^\nu g_{\mu\nu},\,\,\,\,\,\,\,\,\,\,\,\,\,\,\,\,\,\,\,
 T^{+}_{=}= D_- \Phi^\mu \d_= \Phi^\nu g_{\mu\nu}
\eeq{curK1a}
\beq
 {\cal J}_+ = D_+\Phi^\mu D_+\Phi^\nu J_{\mu\nu},\,\,\,\,\,\,\,\,\,\,\,\,\,\,\,\,\,\,\,
 {\cal J}_-=D_-\Phi^\mu D_-\Phi^\nu J_{\mu\nu}.
\eeq{curK1c}
 Using that the manifold is K\"ahler together with the equations of motion
 we derive the following conservation laws
\beq
 D_+ T^{+}_{=}=0,\,\,\,\,\,\,\,\,\,\,\,\,
 D_- T^{-}_{\+}=0,\,\,\,\,\,\,\,\,\,\,\,\,
 D_-J_+=0,\,\,\,\,\,\,\,\,\,\,\,\,
 D_+J_-=0 .
\eeq{conservKc}
 The relevant components of the currents are
\beq
 T_{++} = -iD_+ T^{-}_\+|= \d_{\+}X^\mu \d_{\+}X^\nu g_{\mu\nu} + i 
\psi^\mu_+ \nabla_- \psi^\nu_{+} g_{\mu\nu} , 
\eeq{curTppK}
\beq
 T_{--}= -iD_- T^+_=|=\d_{=}X^\mu \d_{=}X^\nu g_{\mu\nu} + 
 i \psi^\mu_- \nabla_{-} \psi^\nu_{-} g_{\mu\nu}
\eeq{curTmmK}
\beq
 G_+^1= T^-_{\+}|=\psi_+^\mu \d_\+ X^\nu g_{\mu\nu},\,\,\,\,\,\,\,\,\,\,\,\,\,\,
 G_-^1= T^+_=| =\psi_-^\mu \d_= X^\nu g_{\mu\nu}
\eeq{curG1K}
\beq
 G_+^2= -\frac{i}{2} D_+ {\cal J}_+|= \psi_+^\mu \d_\+ X^\nu J_{\mu\nu},\,\,\,\,\,\,\,\,\,\,\,\,\,\,
 G_-^2 = -\frac{i}{2}D_-{\cal J}_-|= \psi_-^\mu \d_= X^\nu J_{\mu\nu}
\eeq{curG2K}
\beq
 J_+={\cal J}_+|=\psi_+^\mu \psi_+^\nu J_{\mu\nu},\,\,\,\,\,\,\,\,\,\,\,\,\,
 J_-={\cal J}_-|=\psi_-^\mu \psi_-^\nu J_{\mu\nu}
\eeq{curJpmK}
 In components the conservation laws  acquire the following form
\beq
\d_{\pp} T_{\mp\mp}=0,\,\,\,\,\,\,\,\,\,\,
\d_{\pp} J_{\mp} =0,\,\,\,\,\,\,\,\,\,\,
\d_{\pp} G^a_{\mp}=0,\,\,\, a=1,2 .
\eeq{conservaKah}

To ensure N=2 superconformal symmetry on the boundary we need to impose the following
 boundary conditions on the currents
\beq
 T_{++}-T_{--}=0,\,\,\,\,\,\,\,\,\,
 G^1_{+}-\eta_1 G^1_{-}=0,\,\,\,\,\,\,\,\,\,
G^2_{+}-\eta_2 G^2_{-}=0,\,\,\,\,\,\,\,\,\,
 J_+ - (\eta_1\eta_2) J_- = 0 .
\eeq{bccurN2K}
 In Section~\ref{s:review} we have solved the first two conditions. Substituting the solutions
 (\ref{formbconshN1}) (where $H=0$) into the last two conditions in (\ref{bccurN2K})  gives
\beq
 J_{\mu\nu}= (\eta_1\eta_2) R^\rho_{\,\,\mu} J_{\rho\sigma} R^\sigma_{\,\,\nu}.
\eeq{jjRRR}
Hence the current analysis coincides exactly with the previous results (cf.(\ref{kahformRR}))

The conserved currents $J_{\pm}$ generate two R-rotations
 which act trivially on the bosonic fields but non-trivially on the fermions.
 Because of the boundary condition $J_{+}-(\eta_1 \eta_2) J_{-}=0$ only one 
 combination of these R-rotations survives as a symmetry in the presence of
 a boundary. Thus for the B-type 
 we have the following R-symmetry
\beq
\left \{ \begin{array}{l}
 \psi_+^\mu \rightarrow \cos \alpha \,\,\psi_+^\mu + \sin \alpha \,\, J^\mu_{\,\,\nu} \psi_+^\nu \\
\psi_-^\mu \rightarrow \cos \alpha \,\,\psi_-^\mu + \sin \alpha \,\, J^\mu_{\,\,\nu} \psi_-^\nu 
\end{array} \right .
 \eeq{RsymBtyK} 
 and for the A-type
\beq
\left \{ \begin{array}{l}
 \psi_+^\mu \rightarrow \cos \alpha \,\,\psi_+^\mu + \sin \alpha \,\, J^\mu_{\,\,\nu} \psi_+^\nu \\
\psi_-^\mu \rightarrow \cos \alpha \,\,\psi_-^\mu - \sin \alpha \,\, J^\mu_{\,\,\nu} \psi_-^\nu 
\end{array} \right . .
 \eeq{RsymAtyK} 
 In complex coordinates these rotations take the familar form: $\psi_{\pm}^i 
 \rightarrow e^{i\alpha}\psi_{\pm}^i$ for the B-type and  $\psi_{\pm}^i 
 \rightarrow e^{\pm i\alpha}\psi_{\pm}^i$ for the A-type. The boundary conditions
 (\ref{holbc}) and (\ref{holbcAAAA}) are invariant under these rotations respectively.

 When using complex coordinates it is sometimes convenient to use a complexified version
 of the currents $G^a_{\pm}$ ($a=1,2$):
\beq
 {\cal G}_{\pm} = \frac{1}{2} (G_{\pm}^1 + i G^2_{\pm}) = \psi_{\pm}^i \d_{\pp} \bar{X}^{\bar{j}}
 g_{i\bar{j}},\,\,\,\,\,\,\,\,\,\,\,\,\,\,\,\,\,\,
\bar{\cal G}_{\pm} = \frac{1}{2} (G_{\pm}^1 - i G^2_{\pm}) = \bar{\psi}_{\pm}^{\bar{i}} \d_{\pp} X^j
 g_{\bar{i}j}
\eeq{complverGK}
 Thus for the B-type case, the complex supercurrent boundary conditions have the form 
\beq
{\cal G}_{+} - \eta {\cal G}_{-} =0,\,\,\,\,\,\,\,\,\,\,\,\,\,\,\,\,\,
\bar{\cal G}_+ -\eta \bar{\cal G}_- =0
\eeq{curBtype}
 Making a R-rotation $\psi_{\pm}^i \rightarrow e ^{\pm i\beta}\psi_{\pm}^i$
 (which is not a symmetry of the model!) we arrive at the generalized B-type boundary conditions
\beq
 {\cal G}_{+} - \eta e^{-2i\beta} {\cal G}_{-} =0,\,\,\,\,\,\,\,\,\,\,\,\,\,\,\,\,\,
\bar{\cal G}_+ -\eta e^{2i\beta}\bar{\cal G}_- =0
\eeq{general}
 ($\eta$ can be absorbed into the phase if one wishes.) Thus these B-type 
 conditions for the currents are solved by the conditions
\beq
 \left \{
\begin{array}{l}
\psi_-^i - e^{-2i\beta} R^i_{\,\,j} \psi^j_{+} = 0\\
\d_= X^i - R^i_{\,\,j}\d_\+ X^j + 4i P^{\bar{k}}_{\,\,\bar{l}} P^i_{\,\,j}
 \nabla_{\bar{k}} Q^j_{\,\,s} \psi_+^s \bar{\psi}_+^{\bar{l}} =0
\end{array} \right .
\eeq{bcnewN2KB}
 where the bosonic boundary condition is related to the fermionic one through the 
 N=2 supersymmetry transformations (\ref{compsusytrN2}) with 
 $\bar{\epsilon}^+ =e^{2i\beta} \bar{\epsilon}^-$.
Although this argument follows the same lines as in \cite{Hori:2000ck}, it should be noted is that 
 the argument is applicable only after the form of the two-fermion terms 
 in the bosonic boundary condition has been established. 

 The same argument can be applied to the A-type boundary conditions. The complex version 
 of the boundary conditions for the supercurrent has the form
\beq
 {\cal G}_+ - e^{-2i\beta} {\bar{\cal G}}_- =0,\,\,\,\,\,\,\,\,\,\,\,\,\,\,\,\,\,
\bar{\cal G}_+ - e^{2i\beta} {\cal G}_- =0 ,
\eeq{bcsupKA}
  and  is solved by 
\beq
 \left \{
\begin{array}{l}
\psi_-^i - e^{-2i\beta} R^i_{\,\,\bar{j}} \bar{\psi}^{\bar{j}}_{+} = 0\\
\d_= X^i - R^i_{\,\,\bar{j}}\d_\+ \bar{X}^{\bar{j}} + i (\nabla_s R^i_{\,\,\bar{j}} +
 R^{\bar{l}}_{\,\,s}\nabla_{\bar{l}} R^i_{\,\,\bar{j}}) \psi_+^s \bar{\psi}_+^{\bar{j}}  =0
\end{array} \right .
\eeq{bcnewN2KA}
 where the bosonic boundary condition is related to the fermionic one through the 
 N=2 supersymmetry transofrmations (\ref{compsusytrN2}) with $\epsilon^+ = e^{2i\beta} \bar{\epsilon}^-$.

 Thus we have established that the conditions (\ref{bcnewN2KB}) and 
 (\ref{bcnewN2KA}) are the most general local solutions of the problem, in the two cases
 (within our framework and given our assumptions).

\subsection{Actions}

 We now briefly review   the derivation of the N=2 superconformal boundary conditions
 starting from an action. There are essentially no new results in this subsection. However 
 hopefully the present discussion will clarify some technical points related to the derivation 
 of the supersymmetric boundary conditions from the action. 

 We start from the action
\beq
 S = \int d^2\sigma\,\left [ \d_\+ X^\mu \d_= X^\nu E_{\mu\nu} + i\psi_+^\mu \nabla_- \psi_+^\nu
 g_{\mu\nu} + i \psi_-^\mu\nabla_+ \psi_-^\nu g_{\mu\nu} + 
 \frac{1}{2}\psi_+^\mu \psi_+^\lambda 
 \psi_-^\rho \psi_-^\nu {\cal R}_{\rho\nu\mu\lambda} \right ] ,
\eeq{actioncomp}
 where $E_{\mu\nu}=g_{\mu\nu}+B_{\mu\nu}$ and it assumed that $dB=0$. 
 The action (\ref{actioncomp}) is the action (\ref{boundBN1act}) written in components with the auxilary 
 field integrated out. 
 The boundary term in the field variation of S is given by 
\beq
 \delta S = i \int d\tau \left [ (\delta\psi_{+}^\mu \psi_+^\nu -
  \delta \psi_{-}^\mu \psi_-^\nu)g_{\mu\nu}  
+ \delta X^\mu ( i
  \partial_\+ X^\nu E_{\nu\mu} - i \partial_= X^\nu E_{\mu\nu} +
  (\psi_-^\nu\psi_-^\rho - \psi_+^\nu \psi_+^\rho
  )\Gamma_{\nu\mu\rho})\right] ,
\eeq{varKBgenf}
 and the supersymmetry variation of S is given by
\beq
\delta_1 S = \epsilon_1^- \int d\tau\,\, \left [ g_{\mu\nu}(\d_\+ X^\mu \psi_-^\nu -
 \eta_1 \d_= X^\mu \psi_+^\nu) + 2\d_0 X^\mu (\psi_-^\nu+\eta_1\psi_+^\nu)B_{\mu\nu} \right ]
\eeq{varn1susy}
 with $\epsilon_1^+=\eta_1\epsilon^1_-$.
Starting from the general fermionic ansatz (\ref{standanR}) wee look for boundary 
 conditions which set both varions (\ref{varKBgenf}) and (\ref{varn1susy}) to zero. These were found
 in \cite{Albertsson:2001dv} and \cite{Albertsson:2002qc}. 
 As discussed above, if the manifold is K\"ahler the action admits 
 an extra supersymmetry.
 The variation of the action (\ref{actioncomp}) under this extra supersymmetry is
 given by 
\beq
\delta_2 S = \epsilon_2^- \int d\tau\,\,\left [ J_{\nu\mu} ( \d_\+ X^\mu \psi_-^\nu -
\eta_2 \d_= X^\mu \psi_+^\nu) + 2\d_0 X^\mu (\psi_-^\lambda + \eta_2 \psi_+^\lambda)
 J^\nu_{\,\,\lambda} B_{\mu\nu} \right ]
\eeq{susyac2}
 with $\epsilon^+_2=\eta_2\epsilon^-_2$. Requiring the variation (\ref{susyac2})
 to vanish we find exactly the same conditions  (\ref{jjRRR}) as  in the previous 
 section. 

 It is convenient to use complex coordinates. We may do 
 this starting from the action (\ref{actioncomp}) and rewriting it in complex form.
  However we must be extra careful if we start from the N=2 action  
\beq
 S= \int d^2\sigma\,d^2\theta\,d^2\bar{\theta}\,\,K(\Phi, \bar{\Phi}) 
\eeq{123AA} 
 and reduce it to the component action in the presence of a boundary.
 The subtlety lies in  the fermionic measure.
 Using (\ref{compspderdef}) and (\ref{spinmesu}) from Appendix B along with the chirality conditions
 on $\Phi$, we obtain  the following expression 
for (\ref{123AA}) reduced to N=1
\beq
S= -\frac{1}{2} \int d^2\sigma\,d^2\theta\,\, \left ( K_{i\bar{j}} D_{[-} \Phi^i
 D_{+]} \bar{\Phi}^{\bar{j}} - D_+ D_- K \right ) 
\eeq{123BB}
 where $K_{i\bar{j}} \equiv \d_i \bar{\d}_{\bar{j}} K$. Comparing (\ref{123BB}) to the 
 standard N=1 action we find an additional boundary term, $D_+ D_- K$. 
 This term is also responsible for the fact that the K\"ahler gauge symmetry 
 corresponding to the transformation
\beq
 K(\Phi, \bar{\Phi})\,\,\,\,\, \rightarrow \,\,\,\,\, K(\Phi, \bar{\Phi}) + h(\Phi) +
 \bar{h} (\bar{\Phi})
\eeq{123CC}   
 is broken in the presence of the boundaries according to
\beq
 \int d^2\sigma\,d^2\theta\,d^2\bar{\theta}\,\,\left [ K(\Phi, \bar{\Phi}) + h(\Phi) +
 \bar{h} (\bar{\Phi}) \right ] =  \int d^2\sigma\,d^2\theta\,d^2\bar{\theta}\,\, 
 K(\Phi, \bar{\Phi}) + \frac{1}{4} \int d\tau\,\,(h_{,i} \d_1 X^i + \bar{h}_{,\bar{i}}
 \d_1 \bar{X}^{\bar{i}}) .
\eeq{123DD} 
 This is very much in analogy to the gauge symmetry for the antisymmetric 2-form 
 gauge field $B$. The gauge symmetry $B \rightarrow B+ d\Lambda$ is also broken by the 
 boundary terms. However that symmetry can be restored by requiring an
 additional tranformation of a $U(1)$ gauge field such that it compensates for the  boundary terms.
 Looking at (\ref{123DD}) it seems reasonable to assume
  that  the boundary terms in (\ref{123DD}) are canceled by
   appropriate transformations of the scalars transverse to the brane
 which couple to $\d_1 X^\mu$ on the boundary.

\subsection{Geometric ineterpretation of the boundary conditions}
\label{s:geometryK}

In this subsection we summarize our results from this section 
 and translate them into statements about the geometry. 

 We have derived the {\em most general local} N=2 superconformal boundary conditions
 in the classical theory. 
 The $N=1$ superconformal boundary conditions correspond to maximal integral submanifolds. 
The extra supersymmetry leads to 
 further restrictions on the submanifolds of the K\"ahler target manifold, ${\cal M}$.
 The B-type boundary condition corresponds to a K\"ahler submanifold, $D$,
 which is invariant under the action of $J$ 
\beq
 J {\cal T}_X(D) \subset {\cal T}_X(D),\,\,\,\,\,\,\,\,\,\,\,\,\,\,\,\,
 J {\cal N}_X(D) \subset {\cal N}_X(D)
\eeq{invarKBt}  
 where ${\cal T}_X(D)$ is the tangent space of $D$ at the point $X$ and
 ${\cal N}_X(D)$ is the normal space at $X$.
 The property (\ref{invarKBt}) follows automatically from (\ref{projBtype}). In fact 
 $J$ induces a complex structure on the submanifold which is 
 integrable and covaraintly constant with respect to the induced 
 connection. Thus the group of the tangent bundle of $D$ can be reduced to $U(k)$ where $\dim(D)=2k$.
 The B-type D-brane may have holomorphic (antiholomorphic)
 gauge fields on it.

 The A-type boundary conditions correspond to the case when the symplectic (K\"ahler) structure 
 resticted to the normal space is zero
\beq
 J|_{{\cal N}_X(D)} = 0,\,\,\,\,\,\,\,\,\,\,\,\,\,\,
 \rank(J|_{{\cal T}_X(D)}) \leq \dim({\cal T}_X(D)) .
\eeq{restricgk22}
Thus A-type branes correspond to  coisotropic manifolds with dimension  
\beq
 \dim(D) = \frac{1}{2}(d +\rank(J|_{{\cal T}_X(D)}) )
\eeq{dimforbr}
 where $d=\dim({\cal M})$ and $\rank(J|_{{\cal T}_X(D)})=\rank(B^\pi)$ which should 
 be muptiple of 4. $B^\pi$ 
 is the field strength of the $U(1)$ field on the brane. The case of zero $B^\pi$
 (i.e., $J|_{{\cal T}_X(D)} = 0$)
 would correspond to a Lagrangian submanifold allowing only flat gauge fields.
 The opposite case of maximal rank of $B^\pi$ correspond to the space-filling brane
 which can be realized for $d=4k$. 
 To have an A-type space-filling brane there must
 be extra geometrical structure on the manifold ${\cal M}$. In the generic situation there is  
 an extra complex structure $\tilde{J}$  with the property (\ref{jjj2222}).
 This situation is realized in, e.g., hyperK\"ahler
 geometry.

 In the generic situation $0 < \rank(J|_{{\cal T}_X(D)}) < \dim({\cal T}_X(D))$
 we have the following decomposition of the tangent space of a D-brane
\beq
 {\cal T}_X(D) = J{\cal N}_X(D) \oplus \hat{\cal T}(D),\,\,\,\,\,\,\,\,\,\,\,\,\,\,
 J{\cal T}_X(D) = {\cal N}_X(D) \oplus \hat{\cal T}(D) .
\eeq{decompsostangent}
We see that the space $\hat{\cal T}(D)$ is invariant under the 
 action of $J$, that is $J\hat{\cal T}(D) =\hat{\cal T}(D)$.
 $J{\cal N}_X(D)$ is the kernel of $J_{\mu\nu}$ restricted to ${\cal T}_X(D)$
 (it is also the kernel of $B^\pi$) and therefore this space is integrable. 
 In the mathematical literature (see, e.g. \cite{Yano2})  submanifolds of this
type
 are sometimes called generic CR submanifolds.
 The complex structures $J$ restricted to the tangent vectors of $D$ 
  gives rise to an f-stucture\footnote{For the definitions and
 the basic properties we refer to the Appendix C} on the submanifold, 
 see equation (\ref{pullfstJ}). In fact, in this case the brane submanifold has
two different 
 f-structures, $J^\pi$ and $\hat{B}^\pi J^\pi$ with  properties (\ref{pullfstJ})
and
 (\ref{pullfstJ11}). These f-srtuctures restricted to $\hat{\cal T}(D)$ become
invertible.
 The presence of the $f$-structure  allows the reduction  of
 the  tangent bundle group of $D$ to $U(2r)\times O(d/2-2r)$ where
$4r=\rank(J^\pi)$. 
 Previously similar observations about the non-Lagrangian A-type branes have
been 
 made in \cite{Kapustin:2001ij}.

 Let us point out a pecularity related to the derivation of the A-type 
 boundary conditions without $B$-field.
A careful look at 
 the previous derivation reveals an essential difference between 
 A- and B-types. Starting from the parity invariant ansatz (\ref{ferman}) with $R^2=I$
 and using the N=2 algebra only would give  
 us two results for the A-type (i.e., $\eta_1\eta_2=-1$);
\beq
 RJR= - J,\,\,\,\,\,\,\,\,\,\,\,\,\,\,\,\,
 P^\mu_{\,\,[\rho} P^\nu_{\,\,\sigma]} \nabla_\mu Q^\lambda_{\,\,\nu} =0
\eeq{algAty123}
 where we recall that $2P=I+R$ and $2Q=I-R$.
 (The derivation is found in subsection~\ref{s:N2algebra}.) For a  geometrical 
 interpretation,  (\ref{algAty123}) is sufficient. The second relation 
 says that we deal with a submanifold and
 the first realtion says that $J$ maps tangent vectors to normal vectors
 and vice versa
\beq
 J {\cal T}_X(D) \subset {\cal N}_X(D),\,\,\,\,\,\,\,\,\,\,\,\,
 J {\cal N}_X(D) \subset {\cal T}_X(D).
\eeq{mapntA} 
 Thus we have a Lagrangian submanifold. Note that this is a purely algebraic result.
 However, to uncover the proper geometrical interpretation for the B-type
 we have to use the current conditions (\ref{bccurN2K}). The condition (\ref{algeintegrablK}) is 
 too week to imply inegrability of $\pi$ and thus we do not obtain a geometrical interpretation 
 from the N=2 algebra alone. This important difference between the A- and B-types is 
 seen  in the full analysis only.  
 
 Further if we want to satisfy property (\ref{deltaXXbound}) for the N=2 superconformal boundary conditions
 we find totally geodesic K\"ahler and Lagrangian submanifolds respectively. 
 Unlike in the N=1 case, this theorem would now hold even in the presence of gauge fields 
 for the B-type boundary conditions.

\Section{N=2 sigma model on bihermitian manifolds} 
\label{s:bihsigma}

We now consider the N=1 superfield bulk action for the real scalar superfields
 $\Phi^\mu$
\beq
 S= \int d^2\sigma\,d^2\theta\,\,D_+\Phi^\mu D_- \Phi^\nu (g_{\mu\nu}(\Phi) 
 + B_{\mu\nu}(\Phi)) ,
\eeq{actionB}
 where we assume that $H\equiv dB\neq 0$. This action is manifestly supersymmetric
 under one supersymmetry because of its N=1 superfield form. Further (\ref{actionB})
 admits an additional nonmanifest supersymmetry
 of the form
\beq
 \delta_2 \Phi^\mu =  \epsilon_2^+ D_+ \Phi^\nu J^\mu_{+\nu}(\Phi)
  + \epsilon_2^- D_- \Phi^\nu J^\mu_{-\nu}(\Phi) 
\eeq{secsupfl}
 where $D_{\pm} \epsilon_2^{\mp}=0$ and  $J^\mu_{\pm\nu}$ are the two complex structures 
 \cite{Gates:1984nk}. 
 The metric $g_{\mu\nu}$ has to be Hermitian
 with respect to both complex structure (hence the name ``bihermitian''). 
 Each of these complex structures is covariantly constant,  with respect 
 to different connections however,
\beq
 \nabla^{(\pm)}_\rho J^\mu_{\pm\nu} \equiv J^\mu_{\pm\nu,\rho} +
 \Gamma^{\pm\mu}_{\,\,\rho\sigma} J^\sigma_{\pm\nu} - \Gamma^{\pm\sigma}_{\,\,\rho\nu}
 J^\mu_{\pm\sigma}=0 ,
\eeq{nablJH}  
 where we defined the two affine connections 
\beq
 \Gamma^{\pm\mu}_{\,\,\rho\nu} = \Gamma^{\mu}_{\,\,\rho\nu} \pm g^{\mu\sigma} H_{\sigma\rho\nu}.
\eeq{defaffcon}
 A few more  formulae are useful.  The inegrability of $J_{\pm}$
  together with (\ref{nablJH}) lead to the following  relation
 \beq
 H_{\delta\nu\lambda} = J^\sigma_{\pm\delta} J^\rho_{\pm\nu} H_{\sigma\rho\lambda} +
 J^\sigma_{\pm\lambda} J^\rho_{\pm\delta} H_{\sigma\rho\nu}+
 J^\sigma_{\pm\nu} J^\rho_{\pm\lambda} H_{\sigma\rho\delta}  .
\eeq{inegrabtroB}
 As another consequence of the constancy of the complex structures we may express
 the torsion $H$ in terms of the complex stuctures $J_{\pm}$, \cite{Gates:1984nk}
\beq
 H_{\mu\nu\rho} = - J^\lambda_{+\mu} J^\sigma_{+\nu} J^\gamma_{+\rho} (dJ_+)_{\lambda\sigma\gamma} = 
 J^\lambda_{-\mu} J^\sigma_{-\nu} J^\gamma_{-\rho} (dJ_-)_{\lambda\sigma\gamma}
\eeq{HtermJplm}
where 
\beq
(dJ_\pm)_{\lambda\sigma\gamma} = \frac{1}{2}(\nabla_\lambda J_{\pm\sigma\gamma} +
\nabla_\sigma J_{\pm\gamma\lambda} + \nabla_\gamma J_{\pm\lambda\sigma}) .
\eeq{defextderJpl}
 The supersymmetry algebra is the usual one. 
 As long as the two complex structures commute,
 the supersymmetry algebra closes off-shell and the model can be formulated in $(2,2)$
 superspace.  
 Commuting complex structures ($[J_- , J_+]=0$) is equivalent to
 the existence of a hermitian locally Riemannian product manifold \cite{Yano2} since 
 then $\Pi = J_-J_+$ is an integrable
 almost product structure. In this case, which is very special from the geometrical 
 point of view, the action (\ref{actionB}) may be rewritten in a manifestly N=2 supersymmetric form
 as
\beq
 S= \int d^2\sigma\,d^2\theta\,d^2\bar{\theta}\,\,K(\Phi, \bar{\Phi}, \Lambda, \bar{\Lambda})
\eeq{bhprodaction} 
 where $\Phi$ and $\Lambda$ are  chiral and twisted chiral 
multiplets respectively. In the case of noncommuting complex structures the supersymmetry 
 algebra closes only on-shell, hence the algebra is model dependent and a manifest supersymmetric 
 formulation will require introduction of additional auxiliary fields. 
 The construction of a manifest 
 off-shell supersymetric version of this model has been investigated in
   \cite{Buscher:uw}-\cite{Bogaerts:1999jc}. 

The main examples of this type of the geometry are given by WZW models. It is also known that 
 any even dimensional group allows for a $N=2$ super Kac-Moody symmetry \cite{Spindel:1988nh}, 
 \cite{Spindel:1988sr}. However for the WZW models
 the metric is never K\"ahler and $H\neq 0$. On these even dimensional group manifolds 
 we find the geometry discussed above realized. 

In what follows we derive formal results for the general case  of a (2,2) sigma model with torsion.
 However due to lack of proper understanding of the underlying geometry it is difficult
 to interpret these results in geometrical terms except for special cases, e.g.
 for commuting complex structures.    

\subsection{The N=2 algebra}
\label{subs:algebrabi}

As in the K\"ahler case a lot of information can be obtained from  algebraic considerations. 
 In components the manifest on-shell supersymmetry transformations are 
\beq
\left \{ \begin{array}{l}
 \delta_1 X^\mu = - (\epsilon^+_1 \psi_+^\mu + \epsilon_1^- \psi_-^\mu) \\
 \delta_1 \psi_+^\mu = -i \epsilon_1^+ \d_\+ X^\mu + \epsilon_1^- 
 \Gamma^{-\mu}_{\,\,\nu\rho} \psi_-^\rho \psi_+^\nu \\
 \delta_1 \psi_-^\mu = -i\epsilon_1^- \d_= X^\mu - \epsilon_1^+ \Gamma^{-\mu}_{\,\,\nu\rho}
 \psi_-^\rho \psi_+^\nu
\end{array}
\right .
\eeq{n11aa}
 and the nonmanifest supersymmetry transformations (\ref{secsupfl}) are 
\beq
\left \{\begin{array}{l}
 \delta_2 X^\mu =   \epsilon^+_2 \psi_+^\nu J^\mu_{+\nu} 
 + \epsilon_2^- \psi_-^\nu J^\mu_{-\nu} \\
 \delta_2 \psi_+^\mu = -i \epsilon_2^+ \d_\+ X^\nu J^\mu_{+\nu} - \epsilon_2^- 
 J^\mu_{-\sigma}\Gamma^{-\sigma}_{\,\,\nu\rho} \psi_-^\rho \psi_+^\nu +
\epsilon_2^+ J^\mu_{+\nu,\rho} \psi_+^\nu \psi_+^\rho + \epsilon_2^- J^\mu_{-\nu,\rho}
 \psi_-^\nu \psi_+^\rho \\
 \delta_2 \psi_-^\mu = -i\epsilon_2^- \d_= X^\nu J^\mu_{-\nu} 
 + \epsilon_2^+ J^\mu_{+\sigma}\Gamma^{-\sigma}_{\,\,\nu\rho}
 \psi_-^\rho \psi_+^\nu +\epsilon_2^+ J^\mu_{+\nu,\rho} \psi_+^\nu \psi_-^\rho
 + \epsilon_2^- J^\mu_{-\nu,\rho}\psi_-^\nu \psi_-^\rho
\end{array}
\right .
\eeq{n11bb}
 As before, we start from the fermionic ansatz (\ref{ferman}) and apply both supersymmetry 
 transformations, (\ref{n11aa}) and (\ref{n11bb}). The result of the first transformation is
\beq
 \d_= X^\mu - R^\mu_{\,\,\nu}\d_\+ X^\nu + 2i (P^\sigma_{\,\,\gamma} \nabla_\sigma
 R^\mu_{\,\,\nu} + P^\mu_{\,\,\rho} g^{\rho\delta} H_{\delta\sigma\gamma}
 R^\sigma_{\,\,\nu})\psi_+^\gamma \psi_+^\nu =0
\eeq{firstsuH}
 where $\epsilon_1^+=\eta_1\epsilon_1^-$.  
 The second supersymmetry gives
\ber
\nonumber
&& \d_= X^\mu + (\eta_1\eta_2) J^\mu_{-\lambda} R^\lambda_{\,\,\sigma} J^\sigma_{+\nu}
\d_\+ X^\nu + i \left [ (\eta_1\eta_2) J^\mu_{-\lambda}\nabla^{(-)}_\rho R^\lambda_{\,\,\nu} 
 J^\rho_{+\gamma}+\right . \\
&&\left . + (\eta_1\eta_2) J^\mu_{-\lambda} R^\lambda_{\,\,\sigma} J^\sigma_{+\rho} H^\rho_{\,\,\nu\gamma}
 + J^\mu_{-\lambda} \nabla^{(+)}_\rho R^\lambda_{\,\,\nu} J^\rho_{-\sigma} R^\sigma_{\,\,\gamma}
 - H^\mu_{\,\,\rho\sigma} R^\sigma_{\,\,\gamma} R^\rho_{\,\,\nu}\right ]\psi_+^\gamma \psi_+^\nu = 0 
\eer{seconsusyH}
 where $\epsilon_2^+=\eta_2\epsilon_2^-$ and we have used the property (\ref{nablJH}). 

 The boundary conditions (\ref{firstsuH}) and (\ref{seconsusyH}) should be equivalent. 
 Starting from the X-part we get the condition
\beq
(\eta_1\eta_2) J^\mu_{-\lambda} R^\lambda_{\,\,\sigma} J^\sigma_{+\nu} = -  
 R^\mu_{\,\,\nu}
\eeq{biherJJRR}
 or equivalently 
\beq
 J^\mu_{-\nu} R^\nu_{\,\,\lambda} =  (\eta_1\eta_2) 
 R^\mu_{\,\,\nu} J_{+\lambda}^\nu .
\eeq{biherjRjR}
 Using $R^\mu_{\,\,\sigma} g_{\mu\nu}R^\nu_{\,\,\rho}
 = g_{\sigma\rho}$ we rewrite (\ref{biherjRjR}) as 
\beq
 R^\mu_{\,\,\sigma} J_{-\mu\nu} R^\nu_{\,\,\rho} = (\eta_1\eta_2) J_{+\sigma\rho} .
\eeq{bihermRRj}
 In analogy with the K\"ahler case we might call the case $\eta_1\eta_2=1$
 a B-type  and $\eta_1\eta_2=-1$ an A-type condition. 
 Using the property (\ref{biherjRjR}) the equation (\ref{seconsusyH}) is rewritten as
\ber
\nonumber
&& 
\d_= X^\mu - R^\mu_{\,\,\nu}
\d_\+ X^\nu + i \left [ (\eta_1\eta_2) J^\mu_{-\lambda}\nabla^{(-)}_\rho R^\lambda_{\,\,\nu} 
 J^\rho_{+\gamma}+\right . \\
&&\left . +
 (\eta_1\eta_2) J^\mu_{-\lambda} \nabla^{(+)}_\rho R^\lambda_{\,\,\nu} R^\rho_{\,\,\sigma} J^\sigma_{+\gamma}
 - R^\mu_{\,\,\lambda} H^\lambda_{\,\,\nu\gamma} 
   - H^\mu_{\,\,\rho\sigma} R^\sigma_{\,\,\gamma} R^\rho_{\,\,\nu}\right ]\psi_+^\gamma \psi_+^\nu = 0 .
\eer{simpBhbbc}
 Using (\ref{inegrabtroB}) we further rewrite (\ref{simpBhbbc}) as 
\beq
 \d_= X^\mu - R^\mu_{\,\,\nu}
\d_\+ X^\nu + 2i J_{+\gamma}^\sigma J_{+\nu}^\lambda \left ( P^\rho_{\,\,\sigma}
 \nabla_\rho R^\mu_{\,\,\lambda} + P^\mu_{\,\,\phi} H^\phi_{\,\,\rho\sigma} R^\rho_{\,\,\lambda}
 \right ) \psi_+^\gamma \psi_+^\nu =0 .
\eeq{simptorHHH}

Comparing the two-fermion terms of (\ref{firstsuH}) and (\ref{simptorHHH}) 
 we get
\beq
 P^\sigma_{\,\,[\gamma|} \nabla_\sigma
 R^\mu_{\,\,|\nu]} + P^\mu_{\,\,\rho}  H^\rho_{\,\,\sigma[\gamma}
 R^\sigma_{\,\,\nu]}
-J_{+[\gamma}^\sigma J_{+\nu]}^\lambda \left ( P^\rho_{\,\,\sigma}
 \nabla_\rho R^\mu_{\,\,\lambda} + P^\mu_{\,\,\phi} H^\phi_{\,\,\rho\sigma} R^\rho_{\,\,\lambda}
 \right ) = 0 .
\eeq{twopferH}
 Using the projectors $\Omega^+_{\pm}=1/2(I\pm iJ_+)$ we rewrite the above 
 condition as
\beq
\Omega_{\pm[\gamma}^{+\sigma} \Omega_{\pm\nu]}^{+\lambda} \left ( P^\rho_{\,\,\sigma}
 \nabla_\rho R^\mu_{\,\,\lambda} + P^\mu_{\,\,\phi} H^\phi_{\,\,\rho\sigma} R^\rho_{\,\,\lambda}
 \right ) = 0 .
\eeq{condomHH1}
 This condition is very similar to the corresponding condition for the K\"ahler case, 
 (\ref{algeintegrablK11}).
 As before  the condition (\ref{condomHH1}) does not imply that the two-fermion term vanishes. 
 However it requires the two-fermion term to have a form which is compatible 
 with an appropriate $U(1)$ R-symmetry.

\subsection{Currents}
\label{subs:curbi}

 Alternatively we derive the superconformal boundary conditions by imposing 
 conditions on the conserved currents. In the form we need them,   
 the currents are
\beq 
 T_{\+}^- = D_{+}\Phi^\mu \partial_{\+} \Phi^\nu g_{\mu\nu} - \frac{i}{3} 
 D_{+}\Phi^\mu D_{+} \Phi^\nu D_{+} \Phi^\rho H_{\mu\nu\rho} , 
\eeq{superc1N2H} 
\beq 
 T_{=}^+ = D_{-}\Phi^\mu \partial_{=} \Phi^\nu g_{\mu\nu} + \frac{i}{3} 
 D_{-}\Phi^\mu D_{-} \Phi^\nu D_{-} \Phi^\rho H_{\mu\nu\rho} , 
\eeq{superc2N2H} 
\beq
 {\cal J}_+ = D_+\Phi^\mu D_+\Phi^\nu J_{+\mu\nu},
\eeq{extracurnN2Hpl}
\beq
  {\cal J}_- = D_-\Phi^\mu D_-\Phi^\nu J_{-\mu\nu}.
\eeq{extracurN2H}
  $T_{\+}^-$  and $T_{=}^+$ correspond to N=1 supercurrent and they were 
 derived for the general case of the N=1 sigma model, e.g. in the Appendix of \cite{Albertsson:2001dv}. 
 The currents
 ${\cal J}_+$ and ${\cal J}_-$ generate the additional supersymmetry transformation (\ref{secsupfl}).
 Using the equations of motion and the properties (\ref{nablJH}) together with the 
 fact the $\nabla^{(\pm)}_\rho
 g_{\mu\nu} = 0$ we find that the currents (\ref{superc1N2H})-(\ref{extracurN2H}) are indeed conserved
\beq 
 D_{+}T_{=}^+ =0,\,\,\,\,\,\,\,\,\,\, 
D_{-}T_{\+}^- =0,\,\,\,\,\,\,\,\,\,\
D_- {\cal J}_+ =0,\,\,\,\,\,\,\,\,\,\
D_+{\cal J}_- =0. 
\eeq{conservlaws} 
The components of the currents (\ref{superc1N2H})-(\ref{extracurN2H}) 
 are related to the (2,2) currents ($T_{\pm\pm}$, $G_{\pm}^1$, $G_{\pm}^2$, $J_{\pm}$)
 as follows
\beq 
T_{++} = -iD_{+} T_{\+}^-| = \d_{\+}X^\mu \d_{\+}X^\nu g_{\mu\nu} + i 
\psi^\mu_+ \nabla^{(+)}_{+} \psi^\nu_{+} g_{\mu\nu} , 
\eeq{comp3} 
\beq 
T_{--} = - iD_{-} T_{=}^+| =  \d_{=}X^\mu \d_{=}X^\nu g_{\mu\nu} + 
 i \psi^\mu_- \nabla^{(-)}_{-} \psi^\nu_{-} g_{\mu\nu} , 
\eeq{comp4N2H}
 \beq 
G^1_{+} = T_{\+}^-| = \psi_{+}^\mu \d_{\+} X^\nu g_{\mu\nu} - \frac{i}{3} 
\psi_{+}^\mu \psi_{+}^\nu \psi_{+}^\rho H_{\mu\nu\rho} , 
\eeq{comp1N2H} 
\beq 
G^1_{-} = T_{=}^+|= \psi_{-}^\mu \d_{=} X^\nu g_{\mu\nu} + \frac{i}{3} 
\psi_{-}^\mu \psi_{-}^\nu \psi_{-}^\rho H_{\mu\nu\rho} , 
\eeq{comp2N2H} 
\beq
 G^2_{+} = \frac{i}{2} D_+ {\cal J}_+| = \psi_+^\mu \d_\+X^\nu J_{+\mu\nu} + 
 \frac{i}{3} \psi_+^\mu \psi^\nu_+ \psi^\rho_+ J^{\lambda}_{+\mu} J^\sigma_{+\nu} J^\gamma_{+\rho}
 H_{\lambda\sigma\gamma}
\eeq{compN2Ha}
\beq
 G^2_{-} =\frac{i}{2} D_- {\cal J}_-| = \psi_-^\mu \d_=X^\nu J_{-\mu\nu} - 
 \frac{i}{3} \psi_-^\mu \psi^\nu_- \psi^\rho_- J^{\lambda}_{-\mu} J^\sigma_{-\nu} J^\gamma_{-\rho}
 H_{\lambda\sigma\gamma}
\eeq{compN2Hb}
\beq
 J_+ = {\cal J}_+| = \psi_+^\mu \psi_+^\nu J_{+\mu\nu},\,\,\,\,\,\,\,\,\,\,\,\,\,\,\,\,
 J_{-} = {\cal J}_-| = \psi_-^\mu \psi_-^\nu J_{-\mu\nu}
\eeq{compN2Hcrsym}
where the covariant derivatives acting on 
the worldsheet fermions are defined by 
\beq 
\nabla^{(+)}_{\pm}\psi_{+}^\nu = \partial_{\pp}\psi_{+}^\nu + 
\Gamma^{+\nu}_{\,\,\rho\sigma}\d_{\pp} X^\rho 
\psi_{+}^\sigma,\,\,\,\,\,\,\,\,\,\, \nabla^{(-)}_{\pm}\psi_{-}^\nu = 
\partial_{\pp}\psi_{-}^\nu + \Gamma^{-\nu}_{\,\,\rho\sigma}\d_{\pp} 
X^\rho \psi_{-}^\sigma . 
\eeq{covdervferm} 
 
To ensure N=2 superconformal symmetry on the boundary 
we impose the following conditions on the 
currents (\ref{comp3})--(\ref{compN2Hcrsym}), 
\beq
 T_{++}-T_{--}=0,\,\,\,\,\,\,\,\,\,
 G^1_+-\eta_1 G^1_- =0,\,\,\,\,\,\,\,\,\,
 G_+^2 - \eta_2 G^2_- =0, \,\,\,\,\,\,\,\,\,
 J_+ - (\eta_1 \eta_2) J_- =0. 
\eeq{curbounN2H}
 The two first conditions were solved completly in \cite{Albertsson:2002qc}. 
 Using these results it is straightforward to find the content of the remaining two conditions. 
 In total, the conditions from \cite{Albertsson:2002qc} has to be supplemented
 by the single condition 
\beq
 R^\rho_{\,\,\mu} J_{-\rho\sigma} R^\sigma_{\,\,\nu} = (\eta_1 \eta_2) J_{+\mu\nu} ,
\eeq{condBIH123} 
 which agrees with the algebraic results from subsection \ref{subs:algebrabi}.

The conserved currents $J_{\pm}$ generate two R-rotations
 which act trivially on the bosonic fields but non-trivially on the fermions.
 Because of the boundary condition $J_{+}-(\eta_1 \eta_2) J_{-}=0$ only one 
 combination of these R-rotations survives as a symmetry in the presence of
 a boundary. Thus for $(\eta_1\eta_2)=1$ 
 we have the following R-symmetry
\beq
\left \{ \begin{array}{l}
 \psi_+^\mu \rightarrow \cos \alpha \,\,\psi_+^\mu + \sin \alpha \,\, J^\mu_{+\nu} \psi_+^\nu \\
\psi_-^\mu \rightarrow \cos \alpha \,\,\psi_-^\mu + \sin \alpha \,\, J^\mu_{-\nu} \psi_-^\nu 
\end{array} \right .
 \eeq{RsymBtyK} 
 and for $(\eta_1\eta_2)=-1$
\beq
\left \{ \begin{array}{l}
 \psi_+^\mu \rightarrow \cos \alpha \,\,\psi_+^\mu + \sin \alpha \,\, J^\mu_{+\nu} \psi_+^\nu \\
\psi_-^\mu \rightarrow \cos \alpha \,\,\psi_-^\mu - \sin \alpha \,\, J^\mu_{-\nu} \psi_-^\nu 
\end{array} \right .
 \eeq{RsymAtyK} 
 Unlike the K\"ahler case these rotations do not simplify in complex form.
  We can choose the complex coordiates with respect to $J_+$ and then the rotations of $\psi_+$
 will takes the simple form: $\psi_+^\mu \rightarrow e^{i\alpha}\psi_+^\mu$. However in these coordinates 
 the rotations of 
 $\psi_-^\mu$ do not have a nice form except in the special case of  commuting
 complex structures. Using the properties (\ref{biherjRjR}) and (\ref{condomHH1}) it is easily checked that 
 the form of the boundary conditions (\ref{ferman}) and (\ref{firstsuH})
  is invariant under a combination of the respective $U(1)$ rotations.

 We face the same problem as above if we try to complexify the supersymmetry 
 currents $G^i_{\pm}$, $i=1,2$. We cannot complexify $G_+^i$ and $G_-^i$ at the same time
 unless the complex structures commute. However we may repeat the same manipulations 
 as for the K\"ahler case in subsection \ref{sub:currents}. As an example, we take 
  $(\eta_1\eta_2)=1$.    
 Making a R-rotation (\ref{RsymAtyK})
 (which is not a symmetry of the model!) we arrive at the generalized 
 $(\eta_1\eta_2)=1$ boundary conditions
\beq
 \left (\begin{array}{l}
 G_+^1 \\
 G_+^2
\end{array} \right ) = \eta \left ( \begin{array}{ll}
 \cos 2\alpha  & -\sin 2\alpha \\
 \sin 2\alpha & \cos 2\alpha
 \end{array} \right )  \left (\begin{array}{l}
 G_-^1 \\
 G_-^2
\end{array} \right )
\eeq{genrbcbi}
 where we used the property (\ref{inegrabtroB}).
 Similar manipulations may be done for the generalized $(\eta_1\eta_2)=-1$ boundary conditions. 

\subsection{Action}

 We derive the N=2 superconformal boundary conditions starting from
 the appropriate action as in \cite{Albertsson:2002qc}. 
  Here we sketch the main steps in  the derivation of  the full set of  
boundary conditions  from the action: The correct action has the form 
\beq 
S= \int d^2\xi d^2\theta\,\, D_{+} \Phi^\mu D_{-} \Phi^\nu E_{\mu\nu} 
(\Phi) -\frac{i}{2}\int d^2\xi\,\,  \partial_{=}( B_{\mu\nu}\psi_{+}^\mu 
\psi_{+}^\nu + B_{\mu\nu} \psi_-^\mu \psi_-^\nu) . 
\eeq{goodaction} 
 or in component 
\beq
 S= \int d^2\xi\,\,\left [ \d_\+ X^\mu \d_= X^\nu E_{\mu\nu} + i \psi^\mu_+\nabla^{(+)}_- \psi_+^\nu
 g_{\mu\nu} + i \psi_-^\mu \nabla^{(-)}_+ \psi^\nu_- g_{\mu\nu} + 
 \frac{1}{2} \psi^\lambda_+ \psi_+^\sigma \psi_-^\rho \psi_-^\gamma 
 {\cal R}^{-}_{\rho\gamma\lambda\sigma} \right ]
\eeq{bihermcomp}
 where we have integrated out the auxilary field.
 In (\ref{bihermcomp}) the curvature is defined as follows
\beq
 {\cal R}^{\pm\mu}_{\,\,\,\sigma\rho\lambda} = \Gamma^{\pm\mu}_{\,\,\,\lambda\sigma,\rho}
 - \Gamma^{\pm\mu}_{\,\,\,\rho\sigma,\lambda} + \Gamma^{\pm\mu}_{\,\,\,\rho\gamma}
 \Gamma^{\pm\gamma}_{\,\,\,\lambda\sigma} - \Gamma^{\pm\mu}_{\,\,\,\lambda\gamma}
 \Gamma^{\pm\gamma}_{\,\,\,\rho\sigma}
\eeq{defcur}
  with $\Gamma^{\pm}$ given by (\ref{defaffcon}) and $\nabla^{(\pm)}_{\mp}$ by (\ref{covdervferm}).
We will show that the result coincides with the conditions obtained 
from the currents in subsection~\ref{subs:curbi}, and hence that the 
above action is indeed the correct one.  

The general field variation of (\ref{bihermcomp}) is
\ber 
\nonumber \delta S &=& i \int d\tau \left [ (\delta \psi_+^\mu \psi_+^\nu - 
\delta \psi_-^\mu \psi_-^\nu)g_{\mu\nu} + \right . \\ 
&& \left .+ \delta X^\mu (i\d_\+ X^\nu 
E_{\nu\mu} - i\d_= X^\nu E_{\mu\nu} + \Gamma^-_{\nu\mu\rho} \psi_-^\nu 
\psi_-^\rho - \Gamma^+_{\nu\mu\rho} \psi_+^\nu \psi_+^\rho) \right ] . 
\eer{bihevarac}
The variation of (\ref{bihermcomp}) under the manifest supersymmetry (\ref{n11aa}) is
\ber 
\nonumber 
\delta_{s1} S &=& \epsilon_1^- \int d\tau \,\,\left [ \d_\+ X^\mu \psi_-^\nu 
E_{\mu\nu} - \eta_1 \psi_+^\mu \d_= X^\nu E_{\mu\nu} + \eta_1 \d_\+ X^\mu 
\psi_+^\nu B_{\mu\nu} + \d_= X^\mu \psi_-^\nu B_{\mu\nu}-\right . \\ 
 && - \left . \frac{i}{3}\eta_1 H_{\mu\nu\rho} \psi_+^\rho \psi_+^\mu 
\psi_+^\nu - \frac{i}{3} H_{\mu\nu\rho} \psi_-^\rho \psi_-^\mu 
\psi_-^\nu \right ] . 
\eer{bihesusy1ac}
The variation  of (\ref{bihermcomp}) under the nonmanifest supersymmetry (\ref{n11bb}) is
\ber 
\nonumber 
\delta_{s2} S &=& \epsilon_2^- \int d\tau \,\,\left [ 
 \eta_2  \d_= X^\mu \psi_+^\rho J^\nu_{+\rho} E_{\nu\mu} 
 - \d_\+ X^\mu \psi_-^\rho J^\nu_{-\rho} E_{\mu\nu}
  - \eta_2 \d_\+ X^\mu \psi_+^\rho J^\nu_{+\rho} B_{\mu\nu} - \right .\\
&& - \left . \d_= X^\mu \psi_-^\rho J^\nu_{-\rho} B_{\mu\nu} 
  +  i J^\lambda_{-\mu} H_{\lambda\nu\rho} \psi_-^\mu \psi_-^\nu 
\psi_-^\rho  + i\eta_2 J^\lambda_{+\mu} H_{\lambda\nu\rho} \psi_+^\mu \psi_+^\nu 
\psi_+^\rho \right ] . 
\eer{bihesusy2ac} 
 Starting from the fermionic ansatz (\ref{ferman}) we have to find  boundary conditions
 that set the three variations (\ref{bihevarac}), (\ref{bihesusy1ac}) and (\ref{bihesusy2ac}) to zero.
 Using the results of \cite{Albertsson:2002qc} we only need to deal with the
  last variation, (\ref{bihesusy2ac}).
 After straightforward calculations we find that the N=1 conditions 
 have to be supplemented by the single condition (\ref{condBIH123}). Thus the action (\ref{bihermcomp}) 
 reproduces the boundary conditions derived from the currents. 

 For commuting complex structures there is an off-shell N=2 represenation of the model in terms
 of chiral and twisted chiral superfields. The action is given by (\ref{bhprodaction}) and there is a gauge
  symmetry for the generalized K\"ahler potential   
\beq
 K (\Phi, \bar{\Phi}, \Lambda, \bar{\Lambda})\,\,\,\,\,\rightarrow\,\,\,\,\,
 K (\Phi, \bar{\Phi}, \Lambda, \bar{\Lambda}) + h(\Phi) +\bar{h}(\bar{\Phi}) + g(\Lambda)
+ \bar{g}(\bar{\Lambda}) .
\eeq{9999AA}
In complete analogy with the K\"ahler case this gauge symmetry is broken by the boundary terms.

\subsection{Geometry}

In the previous subsection we analysed the formal aspects of the N=2
 superconformal boundary conditions for the bihermitian case and found 
 the most general local N=2 superconformal boundary conditions. Following the 
 K\"ahler case we would like to find a geometrical interpretation of these
 boundary conditions in terms of special types of submanifolds. Unfortunately 
 now we cannot use the results from symplecic geometry which 
  we  used
 in the analysis of the K\"ahler case.
 For a N=2 sigma models with  torsion the target manifold is 
 a bihermitian manifold with the additional property (\ref{HtermJplm}). As far as we are aware 
 there is no clear geometrical interpretation of the non-linear condition (\ref{HtermJplm}) which relates
 two complex structures. Therefore we cannot give a full interpretation of the most general solution 
 of the model in simple geometrical terms. 
 Nevertheless we present some partial results on the geometry of these boundary 
 conditions. 

In contrast to the K\"ahler case
 the corresponding submanifold cannot be invariant (anti-invariant) with respecet to one 
 of the complex structures. To illustrate this point, let us assume that 
 a submanifold is $J_-$-invaraint (i.e., $[R, J_-]=0$). We then
 have the following
\beq
 (\eta_1\eta_2) RJ_+ = J_- R= RJ_-,\,\,\,\,\,\rightarrow\,\,\,\,\,R(J_- -(\eta_1\eta_2)J_+)=0 .
\eeq{compman122}
 Since by the definition $R$ is nondegenerate we run into the contradiction that $J_+ = \pm J_-$.  
  Thus $R$ cannot carry purely holomorphic (antiholomorphic) indices with respect to one 
 of the complex structures ($J_+$, e.g).
  
 The next property which is very special for the present model is the relation between the torsion, $H$ 
 and the boundary condition, $R$. As we have discussed there is a non trivial relation 
 (\ref{inegrabtroB})  between  $H$ and $J_{\pm}$ which follows from the inegrability of $J_{\pm}$ and
 the covariant constancy of $J_{\pm}$ with respect to the affine connections with  torsion.  
 In the complex coordinates for $J_+$ ($J_-$) the condition (\ref{inegrabtroB}) have a simple form:
 $H_{ijk}=0$ and $H_{\bar{i}\bar{j}\bar{k}}=0$.  
 Combining the property (\ref{biherjRjR}) with (\ref{inegrabtroB}) we obtain a condition
  which involves $R$
\beq
\left \{ \begin{array}{c}
H^R_{\delta\nu\lambda} = J^\sigma_{+\delta} J^\rho_{+\nu} H^R_{\sigma\rho\lambda} +
 J^\sigma_{+\lambda} J^\rho_{+\delta} H^R_{\sigma\rho\nu}+
 J^\sigma_{+\nu} J^\rho_{+\lambda} H^R_{\sigma\rho\delta} \\
H_R^{\delta\nu\lambda} = J_{-\sigma}^{\delta} J_{-\rho}^{\nu} H_R^{\sigma\rho\lambda} +
 J_{-\sigma}^{\lambda} J_{-\rho}^{\delta} H_R^{\sigma\rho\nu}+
 J_{-\sigma}^{\nu} J_{-\rho}^{\lambda} H_R^{\sigma\rho\delta}
\end{array} \right .  
\eeq{HRHRHRH}
 with the following notation
\beq
 H^R_{\delta\nu\lambda} \equiv R^\mu_{\,\,\delta} R^\rho_{\,\,\nu} R^\sigma_{\,\,\lambda}
 H_{\mu\rho\sigma},
\,\,\,\,\,\,\,\,\,\,\,\,\,\,\,\,\,\,\,
H_R^{\delta\nu\lambda}\equiv  R_{\,\,\mu}^{\delta} R_{\,\,\rho}^{\nu} R_{\,\,\sigma}^{\lambda}
 H^{\mu\rho\sigma} .
\eeq{notHR}
 In canonical coordinates for $J_+$ the property (\ref{HRHRHRH}) has the relatively simple form
\beq
  R^\mu_{\,\,i} R^\rho_{\,\,j} R^\sigma_{\,\,k}
 H_{\mu\rho\sigma} = 0,\,\,\,\,\,\,\,\,\,\,\,\,\,\,\,\,\,
  R^\mu_{\,\,\bar{i}} R^\rho_{\,\,\bar{j}} R^\sigma_{\,\,\bar{k}}
 H_{\mu\rho\sigma} =0
\eeq{simpfadj}
 where $i, j, k$ ($\bar{i}, \bar{j}, \bar{k}$) are
holomorphic indices and $\mu=(i,\bar{i})$. Since $R$ can neither commute nor anticommute 
 with $J_+$  the relation (\ref{simpfadj})
 is a nontrivial restriction on possible $R$'s.

Another common property of these branes is that 
\beq
 Q(J_+ - (\eta_1\eta_2) J_-) Q =0 
\eeq{geomBIQQJJ}
 which follows from (\ref{biherjRjR}) and the property $RQ= QR = -Q$. In geometrical terms
 the property (\ref{geomBIQQJJ}) says that a linear combination of two forms $J_{\pm\mu\nu}$
 is zero when restriced to the normal space
\beq
 (J_{+\mu\nu} - (\eta_1\eta_2) J_{-\mu\nu})|_{{\cal N}_X(D)}=0 .
\eeq{gemBINSJJ}  

 Next we look at the special subclass of the boundary conditions
 with the property that $R^2=I$. In this case it is convenient to introduce
  two (1,1) tensors
\beq
 L^\mu_{\,\,\nu} \equiv J^\mu_{+\nu}+ J^\mu_{-\nu},\,\,\,\,\,\,\,\,\,\,\,\,\,\,\,\,\,
 M^\mu_{\,\,\nu} \equiv J^\mu_{+\nu} - J^\mu_{-\nu}
\eeq{defLM}
 which may be degenerate in a generic situation. 
 Using these tensors we rewrite the condition (\ref{biherjRjR}) as
\beq
 R^\mu_{\,\,\rho} L^\rho_{\,\,\sigma} R^\sigma_{\,\,\nu} = (\eta_1\eta_2) L^\mu_{\,\,\nu},
\,\,\,\,\,\,\,\,\,\,\,\,\,\,\,\,\,\,\,\,
 R^\mu_{\,\,\rho} M^\rho_{\,\,\sigma} R^\sigma_{\,\,\nu} = - (\eta_1\eta_2) M^\mu_{\,\,\nu} .
\eeq{condreLM}  
 Alternatively since the metric is hermitian with respect to both complex structures there 
 exist two corresponding two forms $M_{\mu\nu}$ and $L_{\mu\nu}$.  In terms of these antisymmetric 
 tensors the condition (\ref{condreLM}) is rewritten as
\beq
 R^\mu_{\,\,\rho} L_{\mu\nu} R^{\nu}_{\,\,\sigma} = (\eta_1\eta_2) L_{\rho\sigma},
\,\,\,\,\,\,\,\,\,\,\,\,\,\,\,\,\,\,\,\,
 R^\mu_{\,\,\rho} M_{\mu\nu} R^{\nu}_{\,\,\sigma} = - (\eta_1\eta_2) M_{\rho\sigma} .
\eeq{condreLM22}  
 We first consider the case when $(\eta_1\eta_2)=1$. The relation (\ref{condreLM})
 takes the form
\beq
  [R, L] =0,\,\,\,\,\,\,\,\,\,\,\,\,\,\,\,\,\,\,
 \{ R, M\} =0 .
\eeq{BtypeLMLM} 
 In  full analogy with the analysis from the Section~\ref{s:Kahlersigma}
  we find that the resulting submanifold $D$
 is an invariant submanifold with respect to $L$ and anti-invariant with respect to $M$, i.e.
\beq
 L {\cal T}_X(D) \subset {\cal T}_X(D),\,\,\,\,\,\,\,\,\,\,\,\,\,\,\,\, 
 L {\cal N}_X (D) \subset {\cal N}_X(D),\\
\eeq{inantiLMS1}
\beq 
M {\cal T}_X(D) \subset {\cal N}_X(D),\,\,\,\,\,\,\,\,\,\,\,\,\,\,\,
 M {\cal N}_X(D) \subset {\cal T}_X(D) .
\eeq{inantiLMS2} 
 For the case $(\eta_1\eta_2)=-1$ the role of $L$ and $M$ is interchanged. Thus the resulting 
 manifold is $L$-anti-invariant and $M$-invariant.

 Alternatively  we may construct other (1,1)-tensors, for example the commutator and anticommutator of
 the complex structures. In this case the condition (\ref{biherjRjR}) together with $R^2=I$ implies
\beq
 [J_-, J_+]R= - R[J_-,J_+],\,\,\,\,\,\,\,\,\,\,\,\,\,\,\,
 \{J_-, J_+\}R = R\{J_-, J_+\} .
\eeq{matrixtor}
Thus the submanifold is anti-invariant with respect to $[J_-, J_+]$ and 
invariant with respect to $\{ J_+, J_- \}$ (unless one of them 
 are zero).

 If $L$  and $M$ are non degenerate, then  $\ker [J_+, J_-] = \emptyset$,
 where 
\beq
 \ker [J_+, J_-] = \ker (J_+ - J_-) \oplus \ker(J_+ + J_-) .
\eeq{decomkern}
 In this case the backgound manifold has to be $4k$ dimensional \cite{Bogaerts:1999jc}.
 It is easy to see that when the antisymmetric tensors $L_{\mu\nu}$ and
 $M_{\mu\nu}$ are non degenerate the conditions (\ref{condreLM22}) do not have a solution
 such that $R^2=I$. In the next subsection we will analyse the opposite case when 
 $[J_+, J_-] =0$ and, as we will see, there are many solutions which satisfy $R^2=I$.
 Therefore we conclude that $\ker [J_+, J_-]$ ``controls'' the solutions with
  $R^2=I$ (i.e., without $B$-field). 

 It is difficult to describe the most general case exhaustively.
 Using $L$ and $M$, the condition (\ref{biherjRjR}) for the case $(\eta_1\eta_2)=1$
  can be rewritten as
\beq
 [R, L] = T,\,\,\,\,\,\,\,\,\,\,\,\,\,\,
 \{ R, M\} = T\,\,\,\,\,\,\,\,\,\,\,\,\,\
\eeq{defgenBITLM}
 where $T$ is an auxilary object. For the case $(\eta_1\eta_2)=-1$ $L$ and $M$ would be interchanged
 in (\ref{defgenBITLM}). 
 The fact that $J_{\pm}$ are complex structures
 implies that 
\beq
 \{ L, M\}=0,\,\,\,\,\,\,\,\,\,\,\,\,\,\,\,\,\,\ L^2 + M^2 = -2I .
\eeq{compstMLprop}
 If we consider the space-filling brane then $R$ is a globally defined $(1,1)$ tensor 
 satisfying the algebra (\ref{defgenBITLM}) and (\ref{compstMLprop}).  
 The existence of this algebra on the target manifold should have  non trivial consequences.
 However, we are not familiar with this type of the structures in the mathematical literature.  

\subsubsection{Locally product manifolds}
\label{subs:LPM}

In this subsection we  consider the special case of commuting 
 comlex structures, 
\beq
[J_+, J_-]^\mu_{\,\,\nu} \equiv J^\mu_{+\rho} J^\rho_{-\nu} - J^\mu_{-\rho} J^\rho_{+\nu}= 0 .
\eeq{commutcsH}
 Then the tensor
\beq
 \Pi^\mu_{\,\,\nu} \equiv J^\mu_{+\rho} J^\rho_{-\nu},
\eeq{comtenPI}
 satisfies
\beq
 \Pi^\mu_{\,\,\rho} \Pi^\rho_{\,\,\nu} = \delta^\mu_{\,\,\nu} .
\eeq{almprdst}
 A tensor satsifying above requirement is known as an almost product structure \cite{Yano2}.
 As a result of the inegrability of $J_{\pm}$ the almost product 
 structure $\Pi$ is also integrable. An inegrable $(1,1)$ tensor with the property
 (\ref{almprdst}) gives rise to a local product structure. 
  Using the property (\ref{HtermJplm}) we see that the submanifolds projected out by
\beq
 \Pi_\pm=\frac{1}{2}(I\pm \Pi)
\eeq{projectorsBI}
 are K\"ahler.
 Thus the geometry locally looks like a product of two K\"ahler manifolds, ${\cal M}_1~\times~{\cal M}_2$.
 However the geometry is not a locally decomposable Riemannian manifold, i.e. the metric on ${\cal M}_1$
 depends on the coordinates of ${\cal M}_2$ and vice versa. An example where this situation is realized 
 is the group manifold, $SU(2) \times U(1)$. However, in this example the local product structure $\Pi$ is 
 not the obvious one.

 If we consider solutions with the property $R^2=I$ then, as result of (\ref{biherjRjR}), 
 we get 
\beq
 [\Pi, R ] =0 .
\eeq{commRPI}
 Thus in this case the submanifold corresponding to a D-brane is $\Pi$-invariant, i.e. 
\beq
\Pi {\cal T}_X(D) \subset {\cal T}_X(D),\,\,\,\,\,\,\,\,\,\,\,\,\,\,\,\,\,
\Pi {\cal N}_X(D) \subset {\cal N}_X(D).
\eeq{invarprod}
 Therefore $\Pi$ induces a local product structure on the brane, $D$. 
 $R$ is decomposable in the following form
\beq
 R = \Pi_+ R \Pi_+ + \Pi_- R \Pi_-\equiv R^{+} + R^{-}.
\eeq{decomRlpm}
 Using the above decomposition the problem can be completly separated into two 
 independent problems for the K\"ahler geometry
\beq
 \left \{ \begin{array}{l}
 R^+ J_+ + (\eta_1\eta_2) R^+ J_+ = 0\\
 R^- J_+ - (\eta_1\eta_2) R^- J_+ = 0
\end{array}\right .
\eeq{decomlpmJR}
 where the first equation is understood on ${\cal M}_1$ and the second on ${\cal M}_2$. 
 Thus for the the case $(\eta_1\eta_2)=1$ we have
\beq
 \{R^+, J_+\} =0,\,\,\,\,\,\,\,\,\,\,\,\,\,\,\,\,\,
 [R^-, J_+] = 0
\eeq{123344}
 and the resulting brane is a local product of a Lagrangian submanifold of ${\cal M}_1$ and 
 a K\"ahler submanifold of ${\cal M}_2$ 
 with respect to $J_+$. When $(\eta_1\eta_2)=-1$ the situation is interchanged, i.e. 
 the resulting brane is  local product of a Lagrangian submanifold of ${\cal M}_2$ and 
 a K\"ahler submanifold of ${\cal M}_1$ 
 with respect to $J_+$.  Thus unlike the case with $\ker[J_+, J_-]=\emptyset$
 there are a lot of solutions with the property $R^2=I$.  

 Next if we look for  more general solutions with $R^2\neq I$ then we still 
 construct them in the same fashion assuming the property (\ref{commRPI}). 
  However there are solutions whith  $R^2\neq I$ which do not obey the condition (\ref{commRPI})
 and thus they cannot be thought of as local products of K\"ahler branes. 
In the general situation we solve the problem explicitly in special coordinates. 
 Coordinates exist in which $\Pi$ and $J_{\pm}$ take  block diagonal forms
\beq
 \Pi^\mu_{\,\,\nu} = \left ( \begin{array}{cccc}
  -\delta^n_{\,\,m} & 0 & 0 & 0 \\
   0 & -\delta^{\bar{n}}_{\,\,\bar{m}} & 0 & 0 \\
    0 & 0 & \delta^i_{\,\,j} & 0 \\
    0 & 0 & 0 & \delta^{\bar{i}}_{\,\,\bar{j}}
\end{array} \right ) ,
\eeq{Piincoor}
\beq
 J^\mu_{-\nu} = \left ( \begin{array} {cccc}
    i\delta^n_{\,\,m} & 0 & 0 & 0 \\
   0 & -i\delta^{\bar{n}}_{\,\,\bar{m}} & 0 & 0 \\
    0 & 0 & i\delta^i_{\,\,j} & 0 \\
    0 & 0 & 0 & -i\delta^{\bar{i}}_{\,\,\bar{j}}  
\end{array} \right ),\,\,\,\,\,\,\,\,\,\,\,\,\,\,\,\,\,
 J^\mu_{+\mu} = \left ( \begin{array} {cccc}
     -i\delta^n_{\,\,m} & 0 & 0 & 0 \\
   0 &  i\delta^{\bar{n}}_{\,\,\bar{m}} & 0 & 0 \\
    0 & 0 & i\delta^i_{\,\,j} & 0 \\
    0 & 0 & 0 & -i\delta^{\bar{i}}_{\,\,\bar{j}}  
\end{array} \right ) .
\eeq{speccJJ}
 Thus the condition (\ref{biherjRjR}) with $(\eta_1\eta_2)=1$ can be solved for $R$ as follows
\beq
 R^\mu_{\,\,\nu} = \left ( \begin{array}{cccc}
 0 & R^n_{\,\,\bar{m}} & R^n_{\,\,j} & 0 \\
  R^{\bar{m}}_{\,\,n} & 0 & 0 & R^{\bar{m}}_{\,\,\bar{j}} \\
 0 & R^i_{\,\,\bar{m}} & R^i_{\,\,j} & 0 \\
  R^{\bar{i}}_{\,\,n} & 0 & 0 & R^{\bar{i}}_{\,\,\bar{j}}
\end{array} \right ) .
\eeq{generR1lpm} 
 We see that $R$ does not have a block diagonal form and in the general
 situation the property (\ref{commRPI}) is not true. 
 There are other condition which restricts $R$ further. 
 For example, the condition which involve the metric $R^t g R = g$. 
 The metric $g$ is bihermitian with 
 respect to both complex structure and  has the special form
\beq
 g_{\mu\nu} = \left ( \begin{array}{llll}
 0 & g_{n\bar{m}} & 0 & 0 \\
 g_{n\bar{m}} & 0 & 0 & 0 \\
 0 & 0 & 0 & g_{i\bar{j}} \\
 0 & 0 & g_{i\bar{j}} & 0 
\end{array} \right ) .
\eeq{matricform111}
 However the condition involving the metric does not restrict $R$ to be block diagonal in 
 the $\Pi$-coordinates. Other properties of the branes would depend on the specific properties
 of the model. 
 A similar analysis can be done for the case $(\eta_1\eta_2)=-1$.  

\Section{N=2 Landau-Ginzburg models} 
\label{s:LGmodel}

In this section we consider the special subclass of boundary conditions for
the massive generalization of the N=2 sigma models, so called N=2 Landau-Ginzburg models.

The N=1 Landau-Ginzburg model is given by the bulk action
\beq
 S= \int d^2\sigma\,d^2\theta\,\,\left [D_+\Phi^\mu D_- \Phi^\nu (g_{\mu\nu}(\Phi) 
 + B_{\mu\nu}(\Phi)) + W(\Phi) \right ] .
\eeq{actionBW}
 Generically the classical model is not conformally invariant because of 
 the presence of the potential $W$.
 However it shares some interesting properties with the sigma model. 

 Boundary conditions for the N=1 Landau-Ginzburg model in a trivial background
   are discussed in \cite{rocek}.
 In particular a boundary potential is found to be necessary for supersymmetry.  
In this section we discuss the N=2 version of the Landau-Ginzburg model. 
 The bulk action is manifestly N=1 supersymmetric. Depending on whether 
 the field strength for $B$ is zero or
 nonzero, the nonmanifest supersymmetry has either the form (\ref{extrasusyK}) or 
 the form (\ref{secsupfl}). Therefore the last term in (\ref{actionBW}) should be supersymmetric by itself. 

We first analyse the K\"ahler case. The variation of the last term in (\ref{actionBW}) 
 with respect to the nonmanifest supersymmetry (\ref{extrasusyK}) is  
\beq
 \delta \int d^2\sigma\,d^2\theta\,\,W(\Phi) = \int d^2\sigma\,d^2\theta\,\, W_{,\mu} J^\mu_{\,\,\nu} 
 \epsilon_2^\alpha D_\alpha \Phi^\nu~.
\eeq{varpoteK}
 We require the integrand to be a total derivative and thus find
\beq
 J^\mu_{\,\,\nu} \d_\mu W = \d_\nu \tilde{W}
\eeq{CRhdim}
 where $\tilde{W}$ is an arbitrary function. This condition is a higher dimensional analog of
 the Cauchy-Riemann equations and thus 
 $W$ and $\tilde{W}$ can be thought of as  the real and imaginary parts of a holomorphic function,
 ${\cal W} = W +i \tilde{W}$. Equation (\ref{CRhdim}) implies the following  integrability 
 conditions
\beq    
 \d_{[\rho} (J^\mu_{\,\,\nu]} \d_\mu W) =0.
\eeq{inegrabcond}
 In two dimensions this condition is the requirement that
the real part of holomorphic function should be harmonic.  The bosonic potential can be written 
 either in terms of $W$ (the real part) or in terms of $\tilde{W}$ (the imaginary part) 
\beq
 V(X) = \frac{1}{4} \d_\mu W \d_\nu W g^{\mu\nu} = \frac{1}{4} \d_\mu \tilde{W} 
 \d_\nu \tilde{W} g^{\mu\nu} 
\eeq{bospotrealform}
 When the taget space manifold is K\"ahler and $W$ satisfies (\ref{inegrabcond}) the action can 
 be rewritten in a manifestly N=2 supersymmetric form
\beq
S= \int d^2\sigma\,d^2\theta\,d^2\bar{\theta}\,\,K(\Phi, \bar{\Phi}) +   
\int d^2\sigma\,d^2\theta\,\,{\cal W}(\Phi) + \int d^2\sigma\,d^2\bar{\theta}\,\,\bar{\cal W}(\bar{\Phi})
\eeq{N2LGK}
 where ${\cal W}$ is the holomorphic prepotential defined above. 

If $dB\neq 0$, the taget manifold should be bihermitian and the nonmanifest supersymmetry is 
 given by (\ref{secsupfl}). As in the K\"ahler case we require the potential 
 term to be supersymmetric by itself. Following the previous line of argument,
we take the supersymmetry
 variation of the potential term to be a total derivative and as result we get
 two conditions
\beq
  J^\mu_{+\nu} \d_\mu W = \d_\nu \tilde{W}_+,\,\,\,\,\,\,\,\,\,\,\,\,\,\,\,\,\,\,
  J^\mu_{-\nu} \d_\mu W = \d_\nu \tilde{W}_-
\eeq{bilohol} 
 where $\tilde{W}_+$ and $\tilde{W}_-$ are arbitrary functions. Equations (\ref{bilohol})
 imply integrability conditions for $W$;
\beq
 \d_{[\rho} (J^\mu_{+\nu]} \d_\mu W) =0,\,\,\,\,\,\,\,\,\,\,\,\,\,\,\,\,\,\,
\d_{[\rho} (J^\mu_{-\nu]} \d_\mu W) =0.
\eeq{biholinegrab}
 Therefore we may form two complex functions, ${\cal W}_+= W +i \tilde{W}_+$ and
 ${\cal W}_-= W + i\tilde{W}_-$ which are both holomorphic but with respect to 
 different complex structures.
If the two complex structures commute we may use a manifest 
 N=2 formalism with chiral (antichiral) and twisted chiral (antichiral) supersfields. 
 
 In what follows we will consider a special class of boundary conditions for
 N=2 Landau-Ginzburg models which admits a nice geometrical interpretation. For the K\"ahler
 case the N=2 boundary conditions were studied in \cite{Hori:2000ck} and \cite{Govindarajan:2000my}
 (for earlier study see \cite{3)}).
  We will reproduce some of the results
 of \cite{Hori:2000ck} (but in a different fashion) and we will comment on them. We do not
 know  of a
 study of N=2 boundary conditions for the bihermitian Landau-Ginzburg model. 
 
\subsection{K\"ahler Landau-Ginzburg model}

 We would like to look at a special subclass of N=2 supersymmetric boundary conditions
 for the K\"ahler Landau-Ginzburg model. The on-shell supersymmetry transformations
 have an extra term related to the potential compared with the sigma model. 
In components  the manifest on-shell supersymmetry transformation are 
\beq
\left \{ \begin{array}{l}
 \delta_1 X^\mu = - (\epsilon^+_1 \psi_+^\mu + \epsilon_1^- \psi_-^\mu) \\
 \delta_1 \psi_+^\mu = -i \epsilon_1^+ \d_\+ X^\mu + \epsilon_1^- 
 \Gamma^{\mu}_{\,\,\nu\rho} \psi_-^\rho \psi_+^\nu + \frac{1}{2}\epsilon_1^- g^{\mu\nu} W_{,\nu}\\
 \delta_1 \psi_-^\mu = -i\epsilon_1^- \d_= X^\mu - \epsilon_1^+ \Gamma^{\mu}_{\,\,\nu\rho}
 \psi_-^\rho \psi_+^\nu -\frac{1}{2} \epsilon_1^+ g^{\mu\nu} W_{,\nu}
\end{array}
\right .
\eeq{1susyKLZ}
 and the nonmanifest ones are
\beq
\left \{\begin{array}{l}
 \delta_2 X^\mu =  (\epsilon^+_2 \psi_+^\nu + \epsilon_2^- \psi_-^\nu) J^\mu_{\,\,\nu} \\
 \delta_2 \psi_+^\mu = -i \epsilon_2^+ \d_\+ X^\nu J^\mu_{\,\,\nu} - \epsilon_2^- 
 J^\mu_{\,\,\sigma}\Gamma^{\sigma}_{\,\,\nu\rho} \psi_-^\rho \psi_+^\nu 
 - \epsilon_2^+ J^\mu_{\,\,\nu,\rho} \psi_+^\rho \psi_+^\nu - \epsilon_2^- J^\mu_{\,\,\nu,\rho}
 \psi_+^\rho \psi_-^\nu  - \frac{1}{2} \epsilon_2^- J^\mu_{\,\,\nu} g^{\nu\rho} W_{,\rho} \\
 \delta_2 \psi_-^\mu = -i\epsilon_2^- \d_= X^\nu J^\mu_{\,\,\nu} 
 +\epsilon_2^+ J^\mu_{\,\,\sigma}\Gamma^{\sigma}_{\,\,\nu\rho}
 \psi_-^\rho \psi_+^\nu -\epsilon_2^+ J^\mu_{\,\,\nu,\rho} \psi_-^\rho \psi_+^\nu
 - \epsilon_2^- J^\mu_{\,\,\nu,\rho}\psi_-^\rho \psi_-^\nu + \frac{1}{2}
 \epsilon_2^+ J^\mu_{\,\,\nu} g^{\nu\rho} W_{,\rho}
\end{array}
\right .
\eeq{2susyKLG}
 As before start from the fermion ansatz
\beq
 \psi_-^\mu = \eta_1 R^\mu_{\,\,\nu} \psi_+^\nu .
\eeq{fermanLG}
 However unlike the classical sigma model for Landau-Ginzburg model
  we cannot argue that this ansatz is the unique local ansatz for the fermions
 since there is a dimensionful coupling. In fact, the characteristic property of (\ref{fermanLG}) 
 is that it is a local ansatz which does not contain a dimensionful parameter. 
 We will study only this type of boundary conditions.

 The first supersymmetry transformation (\ref{1susyKLZ}) applied to (\ref{fermanLG})
 yields
\beq
 \d_= X^\mu - R^\mu_{\,\,\nu} \d_\+ X^\nu   + 2i   
 P^\rho_{\,\,\gamma} \nabla_\rho R^\mu_{\,\,\nu}  \psi_+^\gamma \psi_+^\nu - i\eta_1 P^\mu_{\,\,\nu}
 g^{\nu\rho} W_{,\rho} =0 
\eeq{firsusyKLG}
 This is then the bosonic N=1 boundary condition. An important point is that
 the condition (\ref{firsusyKLG}) still has a geometrical inerpretation in terms of submanifolds
  since the last  term in (\ref{firsusyKLG}) does not affect the derivation of the inegrability conditions
 (\ref{inegrabl11}).

 From the second supersymmetry variation of (\ref{fermanLG}) we get
$$ \d_= X^\mu + (\eta_1\eta_2)J^\mu_{\,\,\sigma} R^\sigma_{\,\,\nu} J^\nu_{\,\,\gamma} 
 \d_\+ X^\gamma + i\left [ (\eta_1\eta_2) J^\mu_{\,\,\sigma}
 \nabla_\rho R^\sigma_{\,\,\nu} J^\rho_{\,\,\gamma} + \right .$$ 
\beq
\left . +  J^\mu_{\,\,\sigma} \nabla_\rho R^\sigma_{\,\,\nu} J^\rho_{\,\,\lambda} 
 R^\lambda_{\,\,\gamma}  \right ] \psi_+^\gamma \psi_+^\nu + i\eta_2 P^\mu_{\,\,\nu}g^{\nu\rho} W_{,\rho}
  =0
\eeq{secsuyKLG}
 The boundary conditions (\ref{firsusyKLG}) and (\ref{secsuyKLG}) should be equivalent. Previously we
 have discussed the corresponding boundary conditions
 for the K\"ahler sigma model.  Now for the Landau-Gizburg model the only difference is 
 the last term in (\ref{firsusyKLG}) and (\ref{secsuyKLG}). Therefore all previous results
 from the sigma model analysis apply here as well, including the geometrical 
 interpretation.
 The only new ingredient is that N=2 supersymmetry requires the following property 
 of the potential on the boundary
\beq
 (\eta_1+\eta_2) P^\mu_{\,\,\nu} g^{\nu\rho} W_{,\rho} = 0 .
\eeq{newWKLG}
 When $\eta_1=\eta_2$ (i.e., $\eta_1\eta_2=1$) this does not vanish identically.
 This corresponds to the B-type condition. Using the B-type property that $PJ=JP$
 we see that the holomorphic prepotential ${\cal W}$ is constant along the B-type submanifold
 since
\beq
 P^\mu_{\,\,\nu} g^{\nu\rho} W_{,\rho} = 0,\,\,\,\,\,\,\,\,\,\,\,\,\,
P^\mu_{\,\,\nu} g^{\nu\rho} \tilde{W}_{,\rho} = 0.
\eeq{potenr111}
 This result completly agrees with  \cite{Hori:2000ck}. 

However for the A-type boundary condition the requirement (\ref{newWKLG}) is automatically 
 satisfied. Thus the last term in the bosonic boundary conditions (\ref{firsusyKLG}) is allowed for
 the A-type supersymmetry. To understand the meaning of this term we  ask 
  about the dynamical nature of these boundary conditions. 
 Using the properties (\ref{propN1susy}) we rewrite 
 the bosonic boundary conditions in an equivalent form which may be   
  derived from the following action
\beq
 S = S_{bulk} + \int d\tau V_b (X),\,\,\,\,\,\,\,\,\,(V_{b,\mu} +i\eta_1 W_{,\mu})|_{{\cal T}_X(D)} =0
\eeq{LGactionsBp}
 where $S_{bulk}$ is the sigma model action (\ref{actioncomp}). 
 Thus (as in the N=1 case \cite{rocek}) there is a boundary potential $V_b$
 which is given by the real part (or the imaginary part, depending on the conventions)
  of the bulk prepotential ${\cal W}$. 
  We  stress 
 that this result is valid for all A-type branes (i.e., even with a $B$-field).

 There is interesting subclass of the above boundary conditions. We can require that the bosonic
 boundary condition does not contain a dimensionful parameter. Thus the boundary potential
 should be constant along the brane, i.e.
\beq
 \pi^\nu_{\,\,\mu} W_{,\nu} =0 .
\eeq{propbpetc}
 We conclude
  that the real (or imaginary) part of the prepotential is constant along  brane, 
\beq
 Re ({\cal W}_{,\mu})|_{{\cal T}_X(D)} =0 .
\eeq{vafaresult} 
 This result agrees completly with \cite{Hori:2000ck} (modulo the conventions 
 related to the supersymmetry transformations). 
 However, the derivation persented here is somewhat different. 

 We stress that we do not discuss the general problem of
 the introduction of a boundary potential for the sigma model. In the absence of
 the bulk potential 
 the introduction of a supersymmetric boundary potential would require  non-locality 
 (i.e., the auxilary fields) on the boundary, see for example \cite{Warner:1995ay} and 
 \cite{Husain:2000bt}. In the our case this non-locality is avoided by the 
 presence of the bulk potential in the on-shell supersymmetry transformations. 

\subsection{Bihermitian Landau-Ginzburg model}

 We can repeat a similar analysis for the boundary conditions of the bihermitian 
 Landau-Ginzburg model. However we face the same problem as for the bihermitian sigma 
 model, namely a lack of geometrical understanding of the whole background and
 consequently of the allowed branes on it. Nevertheless below we go through the formal analysis
 and for the case of commuting structure we interpret the result. 

 The manifets on-shell supersymmetry transformations
  are
\beq
\left \{ \begin{array}{l}
 \delta_1 X^\mu = - (\epsilon^+_1 \psi_+^\mu + \epsilon_1^- \psi_-^\mu) \\
 \delta_1 \psi_+^\mu = -i \epsilon_1^+ \d_\+ X^\mu + \epsilon_1^- 
 \Gamma^{-\mu}_{\,\,\nu\rho} \psi_-^\rho \psi_+^\nu +\frac{1}{2} \epsilon_1^- g^{\mu\nu} W_{,\nu}\\
 \delta_1 \psi_-^\mu = -i\epsilon_1^- \d_= X^\mu - \epsilon_1^+ \Gamma^{-\mu}_{\,\,\nu\rho}
 \psi_-^\rho \psi_+^\nu -\frac{1}{2} \epsilon_1^+ g^{\mu\nu} W_{,\nu}
\end{array}
\right .
\eeq{n11aaLG}
 and the nonmanifest supersymmetry transformations (\ref{secsupfl}) are 
\beq
\left \{\begin{array}{l}
 \delta_2 X^\mu =   \epsilon^+_2 \psi_+^\nu J^\mu_{+\nu} 
 + \epsilon_2^- \psi_-^\nu J^\mu_{-\nu} \\
 \delta_2 \psi_+^\mu = -i \epsilon_2^+ \d_\+ X^\nu J^\mu_{+\nu} - \epsilon_2^- 
 J^\mu_{-\sigma}\Gamma^{-\sigma}_{\,\,\nu\rho} \psi_-^\rho \psi_+^\nu +
\epsilon_2^+ J^\mu_{+\nu,\rho} \psi_+^\nu \psi_+^\rho + \epsilon_2^- J^\mu_{-\nu,\rho}
 \psi_-^\nu \psi_+^\rho - \frac{1}{2} \epsilon_2^- J^\mu_{-\sigma} g^{\sigma\lambda} W_{,\lambda}\\
 \delta_2 \psi_-^\mu = -i\epsilon_2^- \d_= X^\nu J^\mu_{-\nu} 
 + \epsilon_2^+ J^\mu_{+\sigma}\Gamma^{-\sigma}_{\,\,\nu\rho}
 \psi_-^\rho \psi_+^\nu +\epsilon_2^+ J^\mu_{+\nu,\rho} \psi_+^\nu \psi_-^\rho
 + \epsilon_2^- J^\mu_{-\nu,\rho}\psi_-^\nu \psi_-^\rho + \frac{1}{2} \epsilon_2^+ J^\mu_{+\sigma}
 g^{\sigma\lambda} W_{,\lambda}
\end{array}
\right .
\eeq{n11bbLG}
 With the analysis in the previous subsection in mind we  derive the special bosonic boundary 
 conditions. Starting from the fermionic ansatz (\ref{fermanLG}) and acting with  the first supersymmetry 
 we find 
\beq
 \d_= X^\mu - R^\mu_{\,\,\nu}\d_\+ X^\nu + 2i (P^\sigma_{\,\,\gamma} \nabla_\sigma
 R^\mu_{\,\,\nu} + P^\mu_{\,\,\rho} g^{\rho\delta} H_{\delta\sigma\gamma}
 R^\sigma_{\,\,\nu})\psi_+^\gamma \psi_+^\nu 
 - i \eta_1 P^\mu_{\,\,\nu} g^{\nu\rho} W_{,\rho} =0
\eeq{firstsuHBIHLG}
 where $\epsilon_1^+=\eta_1\epsilon_1^-$.  
 The second supersymmetry gives
\beq
\d_= X^\mu + (\eta_1\eta_2) J^\mu_{-\lambda} R^\lambda_{\,\,\sigma} J^\sigma_{+\nu}
\d_\+ X^\nu + ...  -\frac{i}{2}  ( \eta_2 J^\mu_{-\lambda} J^\lambda_{+\sigma} + 
 \eta_1 J^\mu_{-\gamma}R^{\gamma}_{\,\,\lambda} J^\lambda_{-\sigma} ) g^{\sigma\rho} W_{,\rho}= 0
\eeq{BILGsecndsusy}
 where the dots stand for the same two fermion term as in equation (\ref{seconsusyH}).
 The boundary conditions (\ref{firstsuHBIHLG}) and (\ref{BILGsecndsusy}) should be equivalent. 
 Previously we have analyzed the $X$-part and two-fermion term for the corresponding N=2 sigma model.
 This analysis extends to the present case of the Landau-Ginzburg model. The only new ingradient is 
 that the N=2 supersymmetry requires the following property of the potential on the boundary 
\beq
  (\delta^\mu_{\,\,\sigma} + R^\mu_{\,\,\sigma} - (\eta_1\eta_2) J^\mu_{-\lambda} J^\lambda_{+\sigma}-
 (\eta_1\eta_2) R^\mu_{\,\,\gamma} J^\gamma_{+\lambda} J^\lambda_{-\sigma} ) g^{\sigma \rho} W_{,\rho}
 = 0
\eeq{newBILG}
 This condition can be understood as a requirement for a brane, $R$, with a given 
 potential, $W$. Following the previous 
 subsection, on the boundary $W$ would be interpreted as a boundary potential. 
  If we restrict ourselves to the case $R^2=I$
 the condition (\ref{newBILG}) can be rewritten in the form
\beq
 ( J^\mu_{-\gamma}+ (\eta_1\eta_2) J^\mu_{+\gamma})
  P^{\gamma}_{\,\,\lambda} g^{\lambda\rho} W_{,\rho} =0 .
\eeq{form1234566}
 where we have used the property (\ref{biherjRjR}).
 We conclude that for branes with $R^2=I$ the structure of $\ker[J_+,J_-]$
 (see equation (\ref{decomkern})) defines the possible restriction on the potential.   

For the case of commuting complex structures the interpretation of the above conditions 
 is clear. The condition (\ref{newBILG}) may be rewritten in the form
\beq
 ( P^\mu_{\,\,\sigma} - (\eta_1\eta_2) P^\mu_{\,\,\lambda} \Pi^\lambda_{\,\,\sigma} )
 g^{\sigma\rho} W_{,\rho} = 0
\eeq{formBJLGcomctsr}
  where $\Pi$ is the product structure from subsection \ref{subs:LPM}. For the case
 of $(\eta_1\eta_2)=1$ boundary conditions we get
\beq
 P^\mu_{\,\,\lambda} \Pi^\lambda_{-\sigma} g^{\sigma\rho} W_{,\rho} =0
\eeq{1234BILG}
 and for the case $(\eta_1\eta_2)=-1$
\beq
 P^\mu_{\,\,\lambda} \Pi^\lambda_{+\sigma} g^{\sigma\rho} W_{,\rho} =0 .
\eeq{1234BILGaa}
  For the composite branes (i.e., $[\Pi, R]=0$) the conditions (\ref{1234BILG})
 and (\ref{1234BILGaa}) say that one has to have a constant $W$ along the B-part 
 of the brane. Thus it agrees completely with the discussion of the K\"ahler case. 
 However the relations (\ref{1234BILG})
 and (\ref{1234BILGaa}) remain true even for the case of non-decomposeable branes. 

In analogy with the K\"ahler case we can consider the situation when the bosonic 
 boundary condition does not contain a dimensionful parameter. 

\Section{Discussion} 
\label{s:discussion} 

 We have presented a
 detailed analysis of the  local superconformal boundary 
 conditions for  N=2 sigma models. Our analysis represents the most general case
 in the sense described in the introduction. Namely, there are various different conditions
 for boundary symmetries that we use, closure of the algebra, gluing of the currents etc.,
 and our results ensure that they are all satisfied. Our starting point was the general 
 local classical condition on the fermions, (\ref{standanR}).

 For the K\"ahler case we reproduce  
 results in a systematic fashion and present an 
 analysis of non-Lagrangian A-branes. Recently non-Lagrangian A-branes have also been  
 discussed in \cite{Kapustin:2001ij}. 

 We stress that in the most 
 general local boundary conditions, the bosonic condition
 has a two fermion term unless the brane is totally geodesic. Since there appears to be some confusion 
 about these issues in the literuture we thought it instructive to go 
 through the details of the general derivation of the boundary conditions for the K\"ahler sigma 
 model. 

Another type of N=2 sigma model has a bihermitian target space with extra 
 properties. This models necessarily involves torsion (a non zero field strength for 
 the NS-NS two form). 
 For bihermitian sigma models we give the full analysis of the local superconformal 
 boundary conditions. Also we discuss the geometrical interpretation of these boundary conditions.   
 
As a natural supplement to the sigma model discussion we consider their massive generalization, 
 the Landau-Ginzburg model. We describe a special subclass of boundary conditions 
 of the Landau-Ginzburg model for which we have a nice geometrical interpretation as 
 submanifolds.

 The main motiovation for our investigation comes from string theory.
 However since our analysis lies entirely in the realm of classical field theory
 and we have tried to maintain a certain level of 
 mathematical rigour,  the present 
 results may be useful for mathematical physics as well.   
 Further, from a string theory point of view the sigma model arise as gauge fixed version
 of the open string action. Correspondingly a complete analysis should also
 take the BRST symmetry for the gauge fixed worldsheet diffeomorphisms into account.
 For certain boundary conditions and trivial background this is done for N=1 in \cite{rocek},
 but there are still open problems in this context.

 One inetersting aspects of our analysis is that the consistent open string 
 sigma models (i.e., the sigma model with boundaries) may require a new geometry on 
 the target space. If we consider freely moving strings (i.e., the space-filling brane)
 we have to introduce a globally defined $(1,1)$ tensor field $R^\mu_{\,\,\nu}$ which  
 encodes the boundary conditions for a sigma model at hand. Requiring  certain 
 symmetries of the model to be preserved in the presence of the boundary amounts to 
 conditions on $R^\mu_{\,\,\nu}$ and therefore  possible to new geometry. 
 For example, the A-type K\"ahler sigma model (i.e., the sigma model with the A-type 
 boundary conditions) would require an extra  complex structure, $\tilde{J}$ with the property
 (\ref{jjj2222}). In the bihermitian case  freely moving 
 strings lead to the structure (\ref{defgenBITLM}) and (\ref{compstMLprop}) which 
 we do not know how to interpret at the moment. We think it worthwhile to persue
 this topic further.
  
 Many open problems remain in the subject. 
 One of them is the search for a better  geometrical understanding of the bihermitian 
 geometry which admits N=2 sigma models with torsion. 
 Such an understanding will shed light on the
 branes which this geometry admits. Another topic 
  would be a 
  detailed analysis (in the spirit of the present work) 
 of the semiclassical branes on Calabi-Yau manifolds. 
 Although there are many things which are known about these (see, e.g., \cite{2)} and \cite{1)}),
  we feel that a rigorous semiclassical analysis would be useful.    
 We hope to come back to these problems elsewhere.

\bigskip 
 
\bigskip

{\bf Acknowledgements}: 
We are grateful Cecilia Albertsson and Martin Ro\v{c}ek for discussions and comments. 
 MZ would like to thank the ITP, Stockholm University and the Department of Theoretical 
 Physics, Uppsala University, where part of
 this work was carried out.
 UL acknowledges support in part by EU contract  
 HPNR-CT-2000-0122 and by VR grant 650-1998368. 
 MZ acknowledges support in part by EU contract HPRN-CT-2002-00325.
 
\appendix 
 
\Section{(1,1) supersymmetry} 
\label{a:11susy} 

 In this and next appendices we present N=1 and N=2 supersymmetries. In our conventions we
 closely follow \cite{Hitchin:1986ea}.
 
We deal with the real (Majorana) two-component spinors $\psi^\alpha= 
 (\psi^+, \psi^-)$. Spinor indices are raised and lowered by the second-rank antisymmetric 
 symbol $C_{\alpha\beta}$, which defines the spinor inner product: 
\beq 
 C_{\alpha\beta}=-C_{\beta\alpha}=-C^{\alpha\beta},\,\,\,\,\, C_{+-}=i,\,\,\,\,\,
\psi_\alpha =\psi^\beta C_{\beta\alpha},\,\,\,\,\, \psi^\alpha= C^{\alpha\beta} \psi_\beta.
\eeq{Cdef} 
Throughout the paper we use  $(\+,=)$ as worldsheet indices, and $(+,-)$ as two-dimensional spinor 
indices.  We also use superspace conventions where the pair of spinor 
coordinates of the two-dimensional superspace are labelled $\th^{\pm}$, 
and the covariant derivatives $D_\pm$ and supersymmetry generators 
$Q_\pm$ satisfy 
\ber 
D^2_+ &=&i\d_\+, \quad 
D^2_- =i\d_= \quad \{D_+,D_-\}=0\cr 
Q_\pm &=& iD_\pm+ 2\th^{\pm}\d_{\pp} 
\eer{alg} 
where $\d_{\pp}=\partial_0\pm\partial_1$.  In terms of the covariant 
derivatives, a supersymmetry transformation of a superfield $\P$ is 
then given by 
\ber 
\delta \P &\equiv & i(\e^+Q_++\e^-Q_-)\P \cr 
&=& -(\e^+D_++\e^-D_-)\P 
+2i(\e^+\th^+\d_\++\e^-\th^-\d_=)\P . 
\eer{tfs} 
The components of a superfield $\P$ are defined via projections as 
follows, 
\ber 
\P|\equiv X, \quad D_\pm\P| \equiv \p_\pm, \quad D_+D_-\P|\equiv F_{+-} 
, 
\eer{comp} 
where a vertical bar denotes ``the $\th =0$ part of ''. 
Thus, in components, the $(1,1)$ supersymmetry transformations are given by 
\beq 
\left \{ \begin{array}{l} 
    \delta X^\mu = - \epsilon^{+} \psi_+^\mu - \epsilon^- \psi_-^\mu \\ 
    \delta \psi_+^\mu =  -i\epsilon^+ \d_{\+}X^\mu + \epsilon^- F^\mu_{+-}\\ 
    \delta \psi_-^\mu  = -i \epsilon^- \d_{=} X^\mu - \epsilon^+ F_{+-}^\mu \\ 
    \delta F^\mu_{+-} = - i \epsilon^+ \d_{\+} \psi_-^\mu + i 
    \epsilon^- \d_- \psi_+^\mu 
\end{array} \right . 
\eeq{compsusytrN1} 
 The N=1 spinorial mesure in terms of covariant derivatives is
\beq
 \int d^2\theta \,\,{\cal L} =   D_+ D_- {\cal L}| .
\eeq{sssssssssspp}

\Section{(2,2) supersymmetry} 
\label{a:11susy} 
 
For N=2 supersymmetry the situation is considerably more involved. Using N=1 formalism
we define complex spinor derivatives
\beq
 D_\alpha\equiv \frac{1}{2}(D_\alpha^1+iD_\alpha^2),\,\,\,\,\,\,\,\,\,\,\,\,\,\,\,
 \bar{D}_\alpha =\frac{1}{2}(D_\alpha^1 - iD^2_\alpha)
\eeq{compspderdef}
 with the algebra
\beq 
\begin{array}{ll} 
 \{D_+, \bar{D}_+ \} = i\d_\+, & 
 \{ D_-, \bar{D}_- \} = i\d_= \\ 
 \{D_\alpha, D_\beta\}= 0,  &\{ \bar{D}_\alpha, \bar{D}_\beta \} =0 .
\end{array} 
\eeq{compder} 
 We also complexify the spinor coordinates. Thus the covariant derivatives
 have the following explicit form
\beq
 D_{\pm} = \d_{\pm}+ \frac{i}{2} \bar{\theta}^{\pm} \d_{\pp},\,\,\,\,\,\,\,\,\,\,\,\,\,\,
 \bar{D}_{\pm} = \bar{\d}_{\pm}+ \frac{i}{2} \theta^{\pm} \d_{\pp}
\eeq{DDbarDD}
In terms of the covariant derivatives, the supersymmetry transformations are
\beq 
 Q_{\alpha}= iD_{\alpha} + \theta^\beta \d_{\alpha\beta}, \,\,\,\,\,\,\,\,\,\,\,\,\,\,
 \bar{Q}_{\alpha} = i\bar{D}_\alpha +\bar{\theta}^\beta \d_{\alpha\beta}
\eeq{QbarQ}
 The supersymmetry transformation of a superfield 
 $\Phi$ is then defined by 
\beq 
 \delta \Phi = i(\epsilon^\alpha Q_\alpha + \bar{\epsilon}^\alpha \bar{Q}_\alpha)\Phi 
\eeq{susyvarD} 
 A chiral superfield ($\bar{D}_{\pm} \Phi =0$) has components 
 defined via projections as follows 
\ber 
\P|\equiv X, \quad D_\pm\P| \equiv \p_\pm, \quad D_+D_-\P|\equiv F_{+-} , 
\eer{compcomp} 
 In components the $(2,2)$ supersymmetry transformations for the chiral 
 multiplet are given by 
\beq 
\left \{ \begin{array}{l} 
    \delta X^i = - \epsilon^{+} \psi_+^i - \epsilon^- \psi_-^i \\ 
    \delta \psi_+^i =  -i\bar{\epsilon}^+ \d_{\+}X^i +  \epsilon^- F^i_{+-}\\ 
    \delta \psi_-^i  = -i \bar{\epsilon}^- \d_{=} X^i - \epsilon^+ F_{+-}^i \\ 
    \delta F^i_{+-} = -i \bar{\epsilon}^+ \d_{\+} \psi_-^i   + i 
    \bar{\epsilon}^- \d_- \psi_+^i 
\end{array} \right . 
\eeq{compsusytrN2} 

The N=2 spinorial mesure in terms of covariant derivatives is
\beq
 \int d^2\theta\,d^2\bar{\theta} \,\,{\cal L} = \left (D_+ D_- \bar{D}_+ \bar{D}_- +
\frac{i}{2} D_+\bar{D}_+ \d_= + \frac{i}{2} D_- \bar{D}_- \d_\+ +
 \frac{1}{4} \d_\+\d_= \right ){\cal L}|
\eeq{spinmesu}
 where the last three terms are purely boundary and thus usually dropped.
 The supersymmetry variation of a general action is
$$ \delta S = i \int d^2\sigma\,d^2\theta\,d^2\bar{\theta} (\epsilon^\alpha Q_\alpha + \bar{\epsilon}^\alpha
 \bar{Q}_\alpha) {\cal L} =$$
\beq
= i  \int d^2\sigma\,\,\left (\epsilon^\alpha D_{\alpha} + 
 \bar{\epsilon}^\alpha \bar{D}_\alpha \right )\left ( D_+ D_- \bar{D}_+ \bar{D}_- +
 \frac{i}{2} D_+\bar{D}_+ \d_= +
\frac{i}{2} D_- \bar{D}_- \d_\+ +
 \frac{1}{4} \d_\+\d_= \right ){\cal L}| 
\eeq{susyvarn2}
 where again we have kept all boundary terms.

\Section{Submanifolds} 
\label{a:submanifolds}

 In this appendix we summarise the relevant mathematical details on 
submanifolds of Riemannian manifolds.  In our use of terminology we 
closely follow \cite{Yano2}. 
 
We first give the definition of a \emph{distribution} on a manifold (or 
neighbourhood) ${\cal M}$.  A distribution $\pi$ of dimension $(p+1)$ 
on ${\cal M}$ is an assignment to each point $X \in {\cal M}$ of a 
$(p+1)$-dimensional subspace $\pi_X$ of the tangent space ${\cal T}_X({\cal 
M})$.  The assignment can be done in different ways, for instance by 
means of an appropriate projection operator.  $\pi$ is called 
\emph{differentiable} if every point $X$ has a neighbourhood $U$ and $(p+1)$ 
differentiable vector fields, which form a basis of $\pi_Y$ at every 
$Y \in U$.  $\pi$ is called \emph{involutive} if for any two vector fields 
$v_i$, $v_j$ $\in \pi_X$ their Lie bracket $\{ v_i, v_j \} \in \pi_X$ 
for all $X \in {\cal M}$. 
 
A connected submanifold $D$ of ${\cal M}$ is called an \emph{integral 
manifold} of $\pi$ if $f_*({\cal T}_X(D))=\pi_X$ for all $X \in D$, where $f$ 
is the embedding of $D$ into ${\cal M}$.  If there is no other 
integral manifold of $\pi$ which contains $D$, then $D$ is called a 
\emph{maximal integral manifold} of $\pi$. 
 
{\bf Frobenius theorem}: Let $\pi$ be an involutive distribution on a 
manifold ${\cal M}$. Then through every point $X \in {\cal M}$, there 
passes a unique maximal integral manifold $D(X)$ of $\pi$.  Any 
other integral manifold through $X$ is an open submanifold of $D(X)$. 
 
 Now we can defined the distribution by means of
 the  projectors. Let $Q^\mu_{\,\,\nu}(X)$ be
a differentiable distribution\footnote{We need to assume 
differentiability to be able to do the calculations.} which assigns to 
a point $X$ in the $d$-dimensional spacetime manifold ${\cal M}$ a 
$(d-p-1)$-dimensional subspace\footnote{We take $\rank(Q)=d-p-1$ in order to match 
 the D-brane terminology.}  of 
the tangent space ${\cal T}_X({\cal M})$.  This subspace consists of all 
vectors $v^\mu(X) \in {\cal T}_X({\cal M})$ such that 
\beq 
 Q^\mu_{\,\,\nu}(X) v^\nu(X) = v^\mu(X) . 
\eeq{Qdef}  
The complementary distribution is defined as 
$\pi^\mu_{\,\,\nu}=\delta^\mu_{\,\,\nu}- Q^\mu_{\,\,\nu}$, 
and assigns to $X$ a $(p+1)$-dimensional space 
that consists of vectors $v^\mu(X) \in T_X({\cal M})$ such that 
\beq 
 \pi^\mu_{\,\,\nu}(X) v^\nu(X) = v^\mu(X). 
\eeq{pidef} 
Now we ask when the vector fields defined by (\ref{pidef}) form a 
submanifold. The Lie bracket of two vector fields $v$ and $w$ 
in $\pi$-space is 
\beq 
\{ v, w\}^\nu = \pi^\nu_{\,\,\sigma} \{ v, w\}^\sigma + 
v^\rho w^\sigma \pi^\mu_{\,\,\sigma} \pi^\lambda_{\,\,\rho} 
Q^\nu_{\,\,[\lambda,\mu]} . 
\eeq{Frobth} 
If the last term vanishes $\pi^\mu_{\,\,\sigma} \pi^\lambda_{\,\,\rho} 
Q^\nu_{\,\,[\lambda,\mu]} = 0$   (i.e., if $\pi^\mu_{\,\,\nu}$ is 
integrable), then the distribution $\pi^\mu_{\,\,\nu}$ 
 is involutive, and due to the 
Frobenius  theorem  there is a unique \emph{maximal 
  integral submanifold} corresponding to $\pi^\mu_{\,\,\nu}$. 
 
If the manifold ${\cal M}$ is Riemannian, then various structures may 
be induced on the submanifold $D$. For instance, $D$ is automatically 
Riemannian. If one defines the Levi-Civita connection 
$\nabla_v \equiv v^\mu \nabla_\mu$ on ${\cal M}$, and takes two 
vector fields $v$ and $w$ in the tangent space ${\cal T}(D)$ of $D$, then 
the covariant derivative $\nabla_v w$ can be decomposed as 
\beq 
 \nabla_v w = 
\hat{\nabla} _v w + {\cal B}(v,w), 
\eeq{decomcov} 
where $\hat{\nabla} 
_v w$ is the tangential component (i.e., it is in ${\cal T}(D)$) and ${\cal 
B}(v,w)$ is the 
normal component. One can show that $\hat{\nabla} _v$ can serve as 
the induced connection on the submanifold $D$. ${\cal B}$ is called the 
\emph{second fundamental form} of $D$. Sometimes it is useful to introduce 
the \emph{associated second fundamental form}, ${\cal A}$, which 
is defined as follows.  Taking $z$ to be 
a normal vector field on $D$ and $v$ a tangent vector field on $D$ we write 
\beq 
 \nabla_v z = -{\cal A}_z v + D_v z  
\eeq{secassff} 
 where $-{\cal 
A}_z v$ and $D_v z$ are, respectively, the tangential and the normal 
components of $\nabla_v z$. Using the metric $g$ on ${\cal M}$ one can 
prove the following simple identity, 
\beq 
 g( {\cal B}(v,w), z)= g({\cal A}_z v, w) . 
\eeq{ABrel} 
Eqs.~(\ref{decomcov}) and 
(\ref{secassff}) are called the Gauss formula and the Weingarten 
formula, respectively. A submanifold $D$ is said to be \emph{totally geodesic} if 
 its second fundamental form vanishes identically, that is, ${\cal B}=0$
 or equivalently ${\cal A}=0$.
 
Now we can rewrite the above definitions in terms of the projectors.  
Take two vector fields $v$, $w$ in the $\pi$-space. 
Denoting by $\nabla$ the connection on ${\cal M}$, we may write 
\beq 
v^\mu \nabla_\mu w^\nu = 
\pi^\nu_{\,\,\rho} v^\mu \nabla_\mu w^\rho 
 + Q^\nu_{\,\,\rho} v^\mu \nabla_\mu w^\rho, 
\eeq{e:nablavdecomp} 
where we decomposed the derivative into its tangential 
(to the $\pi$-space) and normal parts 
by using $\delta^\mu_{\,\,\nu} = \pi^\mu_{\,\,\nu} + Q^\mu_{\,\,\nu}$. 
The tangential component is the induced connection, and 
the normal component is the second fundamental form (\ref{decomcov}). 
The latter may be rewritten, 
using that $\pi^\mu_{\,\,\nu}v^\nu = v^\mu$ and 
$\pi^\mu_{\,\,\nu}w^\nu = w^\mu$, as 
\beq 
v^\delta w^\sigma \, {\cal B}^\lambda_{\,\,\delta\sigma} 
\equiv -v^\delta w^\sigma \,\, \pi^\mu_{\,\,\delta} \pi^\nu_{\,\,\sigma} 
\nabla_\mu Q^\lambda_{\,\,\nu}. 
\eeq{secff} 
Note that ${\cal B}^\lambda_{\,\,\delta\sigma}$ is symmetric 
in the indices $\delta$ and $\sigma$, as a second fundamental form must be, 
because $\pi^\mu_{\,\,\nu}$ is integrable. 
 
Performing the same decomposition for the derivative of 
a vector field $u$ in the $Q$-space, we have 
\beq 
v^\mu \nabla_\mu u^\nu = 
\pi^\nu_{\,\,\rho} v^\mu \nabla_\mu u^\rho 
 + Q^\nu_{\,\,\rho} v^\mu \nabla_\mu u^\rho, 
\eeq{e:nablawdecomp} 
where $Q^\mu_{\,\,\nu} u^\nu = u^\mu$ and $v$ is in 
the $\pi$-space, $\pi^\mu_{\,\,\nu}v^\nu = v^\mu$. 
The associated second fundamental form is then defined as the 
tangential part, which we can rewrite as 
\beq 
v^\delta u^\sigma \, {\cal A}^\lambda_{\,\,\sigma\delta} 
 \equiv - v^\delta u^\sigma \,\, \pi^\mu_{\,\,\delta} 
\pi^\lambda_{\,\,\nu} \nabla_\mu Q^\nu_{\,\,\sigma} . 
\eeq{asff} 
Thus the manifold is totally geodesic if and only if one of the equivalent 
 properties holds
\beq
\pi^\mu_{\,\,\delta} 
\pi^\lambda_{\,\,\nu} \nabla_\mu Q^\nu_{\,\,\sigma} =0\,\,\,\,\,\,\,\,\,\,
or\,\,\,\,\,\,\,\,\,\, \pi^\mu_{\,\,\delta} \pi^\nu_{\,\,\sigma} 
\nabla_\mu Q^\lambda_{\,\,\nu} =0
\eeq{totgeodKapp} 

If on a manifold ${\cal M}$ there is $(1,1)$ non-null tensor field, $L$ then 
 one can consider the invariant submanifolds under the action of $L$, i.e. 
\beq
 L{\cal T}_X(D) \subset {\cal T}_X(D),\,\,\,\,\,\,\,\,\,\,\,\,\,\,\,\,
L{\cal N}_X(D) \subset {\cal N}_X(D)
\eeq{definKahlman}
 and anti-invariant submanifolds under the action $L$, i.e.
\beq
 L{\cal T}_X(D) \subset {\cal N}_X(D),\,\,\,\,\,\,\,\,\,\,\,\,\,\,\,\,
L{\cal N}_X(D) \subset {\cal T}_X(D)
\eeq{definKahlmanAAA}

\Section{Symplectic Geometry} 
\label{a:symplectic} 

 A manifold is called \emph{symplectic} if there exists a nondegenerate
 closed two-form $\omega$. This two-form is a symplectic form. 
 In order to classify the submanifold of a symplectic manifold
 one has to classify the subspaces of the symplectic linear space.
 
Let $W$ be a $k$-dimensional subspace of the $2n$-dimensional symplectic 
 space $(V, \omega)$. Then $k=dim(W)$  and $2l=rank(\omega|_W)$ remain unchanged
 under any symplectic morphism from $Sp(V)$. Therefore these two integers, $k$ and
 $2l$, classify the subspaces of $V$ and they are the only two independent symplectic invariants
 for subspaces. The most important cases of subspaces of a symplectic space $(V,\omega)$
 are the following.\\
\,* A subspace $W\subset V$ with $\omega|_W$ non-degenerate is called 
 a \emph{symplectic subspace}. \\
\,* A subspace $W \subset V$ with $\omega|_W=0$ is called an \emph{isotropic subspace}.\\
\,* A subspace $W \subset V$ with $W^\perp$ isotropic is called  \emph{coisotropic}.\\
\,* A subspace $W\subset V$ which is both isotropic and coisotropic is called 
 a  \emph{Lagrangian subspace}. 
In the above definitions $W^\perp$ is the $\omega$-orthogonal space
$$ W^\perp \equiv \{ v \in V; \omega(v,w)=0, \forall w \in W \}$$ 
 Alternatively, one can give the follwing definitions for a subspace $W \subset V$
 with $k=dim(W)$
 $$\,W\,\,\,\,\,isotropic\,\,\,\,\,\Leftrightarrow\,\,\,\,\,W \subset W^\perp\,\,\,\,\,
\rightarrow\,\,k \leq n$$
 $$\,W\,\,\,\,\,coisotropic\,\,\,\,\,\Leftrightarrow\,\,\,\,\,W \supset W^\perp\,\,\,\,\,
\rightarrow\,\,k \geq n$$
 $$\,W\,\,\,\,\,Lagrange\,\,\,\,\,\Leftrightarrow\,\,\,\,\,W = W^\perp\,\,\,\,\,
\rightarrow\,\,k = n$$
 When this classification is applied to the tangent space ${\cal T}(D)$ of a submanifold $D$
 of a symplectic manifold ${\cal M}$ we have the classification of 
 submanifolds of a symplectic manifold.

 For a general introduction to symplectic geometry the reader can consult, for example,  
 the following book \cite{symlectic}.

\Section{f-structures} 
\label{a:fstructures} 

A structure on an $d$-dimensional manifold ${\cal M}$ given by a non-null tensor
 field $f$ satisfying
\beq
   f^3 + f = 0
\eeq{definfstr}
 is called an $f$-structure (due to Yano). Then the rank of $f$ is constant.
 If $d=\rank(f)$, then an $f$-structure gives an almost complex structure (i.e., $f^2=-I$)
 of the manifold ${\cal M}$. If ${\cal M}$ is orientable and $d-1=\rank(f)$, then an $f$-structure
 gives an almost contact structure of the manifold ${\cal M}$. 
 An $f$-stucture is called integrable if the Nijenhuis tensor of $f$ is zero. 

{\bf Theorem}: A necesary and sufficient condition for a 
 $d$-dimensional manifold ${\cal M}$ to admit an $f$-structure $f$ is that the rank
  of $f$ is even, $\rank(f)=2m$, and that the group of the tangent bundle of ${\cal M}$ be reduced
 to the group $U(m)\times O(d-2m)$.

\end{document}